\documentclass[a4paper,11pt]{article}
\pdfoutput=1 

\usepackage{jcappub} 

\usepackage[T1]{fontenc} 
\usepackage{placeins}
\usepackage[dvipsnames,svgnames,x11names,hyperref]{xcolor}
\usepackage{array}
\usepackage{graphicx}
\usepackage{booktabs}
\usepackage[normalem]{ulem}
\usepackage{cancel}
\usepackage{slashed} 
\RequirePackage{mathrsfs}
\usepackage{dsfont}
\usepackage{hyperref}
\usepackage{subcaption}
\usepackage{amsmath}
\hypersetup{
        unicode = true,
        pdftitle = Note , 
        colorlinks = true,
        linkcolor = Black!30!DodgerBlue4,
        citecolor = Black!30!DodgerBlue4,
        filecolor = Black!30!DodgerBlue4,
        urlcolor = Black!30!DodgerBlue4,
}

\usepackage{booktabs}
\usepackage{cancel}

\graphicspath{{./plots/}}

\usepackage{floatrow}
\newfloatcommand{capbtabbox}{table}[][\FBwidth]

\usepackage{ragged2e}

\renewcommand{\epsilon}{\varepsilon}

\renewcommand{\vec}[1]{\boldsymbol{#1}}
\renewcommand{\bar}{\overline}
\renewcommand{\Re}{\mathrm{Re}}
\renewcommand{\Im}{\mathrm{Im}}

\newcolumntype{L}{>{$}l<{$}}
\newcolumntype{C}{>{$}c<{$}}

\allowdisplaybreaks

\title{\boldmath \justifying Spin-1 Thermal Targets for Dark Matter Searches at Beam Dump and Fixed Target Experiments}

\author[a]{Riccardo Catena}
\author[a]{and Taylor R. Gray}

\affiliation[a]{Chalmers University of Technology, Department of Physics, SE-412 96 G\"oteborg, Sweden}

\emailAdd{catena@chalmers.se}
\emailAdd{taylor.gray@chalmers.se}

\abstract{The current framework for dark matter (DM) searches at beam dump and fixed target experiments primarily relies on four benchmark models, the so-called complex scalar, inelastic scalar, pseudo-Dirac and finally, Majorana DM models.~While this approach has so far been successful in the interpretation of the available data, it a priori excludes the possibility that DM is made of spin-1 particles -- a restriction which is neither theoretically nor experimentally justified.~In this work we extend the current landscape of sub-GeV DM models to a set of models for spin-1 DM, including a family of simplified models (involving one DM candidate and one mediator -- the dark photon) and an ultraviolet complete model based on a non-abelian gauge group where DM is a spin-1 Strongly Interacting Massive Particle (SIMP).~For each of these models, we calculate the DM relic density, the expected number of signal events at beam dump experiments such as LSND and MiniBooNE, the rate of energy injection in the early universe thermal bath and in the Intergalactic Medium (IGM), as well as the helicity amplitudes for forward processes subject to the unitary bound.~We then compare these predictions with experimental results from Planck, CMB surveys, IGM temperature observations, LSND, MiniBooNE, NA64, and BaBar and with available projections from LDMX and Belle II.~Through this comparison, we identify the regions in the parameter space of the models considered in this work where DM is simultaneously thermally produced, compatible with present observations, and within reach at Belle II and, in particular, at LDMX.~We find that the simplified models considered here are strongly constrained by current beam dump experiments and the unitarity bound, and will thus be conclusively probed (i.e.~discovered or ruled out) in the first stages of LDMX data taking.~We also find that the vector SIMP model explored in this work predicts the observed DM relic abundance, is compatible with current observations and within reach at LDMX in a wide region of the parameter space of the theory. } 

\begin{document} 
\maketitle
\flushbottom

\section{Introduction}
The lack of discovery of Weakly Interacting Massive Particles (WIMPs) at dark matter (DM) direct detection experiments has motivated the exploration of a variety of alternative theoretical and experimental paradigms over the past decade~\cite{Cooley:2022ufh,Battaglieri:2017aum}. In this exploration, emphasis has been placed on probing DM candidates lying outside the canonical WIMP mass window, with most of these efforts focusing on the MeV -- GeV mass range~\cite{Mitridate:2022tnv}. This choice is supported by at least three reasons~\cite{Essig:2011nj}. First, a DM candidate lighter than a nucleon would not carry enough kinetic energy to induce an observable nuclear recoil in a direct detection experiment, thereby explaining in a simple and economical way the lack of discovery of WIMPs. Second, the present cosmological density of particles in this mass range can match the one observed for DM by the Planck satellite~\cite{Planck:2018vyg}.~This can occur via the chemical decoupling mechanism if the new sub-GeV states have interactions that involve new particle mediators in the same mass range, thus evading the Lee-Weinberg bound~\cite{Lee:1977ua}. Finally, the sub-GeV DM hypothesis can be tested experimentally using existing methods, including direct detection experiments sensitive to DM-induced electronic excitations in materials, as well as beam dump and fixed target experiments. Especially important for this work are the operating beam dump experiments LSND~\cite{LSND:2001akn} and MiniBooNE~\cite{MiniBooNEDM:2018cxm}, and the fixed target experiment LDMX~\cite{Akesson:2018vlm}, as they can effectively probe sub-GeV DM models with DM-electron or -nucleon scattering cross sections that are suppressed by small DM velocities or momentum transfers.

The theoretical framework currently used in the analysis of operating beam dump experiments, as well as in assessing the prospects of next-generation beam dump and fixed target experiments consists of four benchmark models~\cite{Berlin_2019,deNiverville_2019,deNiverville_2017,deNiverville:2016rqh,E137,deFavereau:2013fsa,deNiverville:2011it}, often referred to as complex scalar DM, scalar inelastic DM, pseudo-Dirac DM, and, finally, Majorana DM.~While reviewing these four models goes beyond the scope of this introduction, it is apparent that this theoretical framework a priori excludes the possibility that DM consists of spin-1 particles.~However, as pointed out by different groups in a series of recent works focusing on the direct detection of vector DM~\cite{Chu:2023zbo,Gondolo:2021fqo,Gondolo:2020wge,Catena:2019hzw,Catena:2018uae}, there is no theoretical or experimental argument supporting this restriction.

The main purpose of this work is to extend the current framework for DM searches at beam dump and fixed target experiments to the case of spin-1 DM.~As a first step towards this extension, we study the phenomenology of a set of simplified models for vector DM featuring a kinetic mixing between an ordinary and a ``dark photon''~\cite{Catena:2022fnk,Dent:2015zpa}.~The latter is responsible for mediating the interactions between DM and the known electrically charged particles.~These simplified models conceptually extend the Standard Model (SM) of particle physics in the same minimal way as the four benchmark models listed above.~Next, we will focus on the phenomenology of an ultraviolet complete model where DM is made of Strongly Interacting Massive Particles (SIMPs)~\cite{Choi:2019zeb}.~In both cases, we identify the regions in the parameter space of the theory where DM can be thermally produced, is not excluded by current experiments and, finally, is within reach at LDMX~\cite{LDMX}.~One usually refers to these regions as {\it thermal targets}.

We find that the simplified models for spin-1 DM that we consider here are subject to strong constraints from existing beam dump experiments, as well as from the unitarity of the S-matrix.~We also find that the regions in the parameter space of these models that are not already ruled out by existing theoretical and experimental constraints will soon be probed at LDMX, which will conclusively discover or exclude this family of models in the early stage of data taking.~In contrast, in the case of vector SIMP DM, we find that the observed DM cosmological density can be reproduced in a broader region of parameter space.~A significant fraction of this is not excluded by existing beam dump experiments, and is within reach at LDMX. 

This article is organised as follows.~We start by introducing the spin-1 DM models explored in our work in Sec.~\ref{sec:theory}.~We then review the experimental and theoretical constraints these models have to fulfill in Sec.~\ref{sec:bounds}.~The main results of our analysis are reported in Sec.~\ref{sec:results}, where we identify the regions of the parameter space of our models in which DM is not ruled out by current experiments, is thermally produced, and is within reach at LDMX.~Finally, we summarise and conclude in Sec.~\ref{sec:conclusion}.~Useful scattering cross section formulae are listed in the appendices.

\section{Models for vector dark matter}
\label{sec:theory}
In this section, we provide a brief review of the spin-1 DM models that we consider in this work. We focus on so-called ``simplified models'' featuring one DM candidate and a single new particle mediator in the mass spectrum, as well as on a renormalisable, ultraviolet complete model where the DM candidate is a Strongly Interacting Massive Particle (SIMP).~The first framework enables us to extend the study of DM at fixed target and beam dump experiments in a way that can directly be compared with the existing literature on scalar and fermionic DM.~As we will see, this first approach is constrained by existing data and subject to strong bounds from the unitarity of the S-matrix, and will therefore be conclusively probed in the early stages of LDMX data taking.~The second framework is by construction compatible with the unitarity of the S-matrix, and it complements our simplified model analysis by focusing on a different mechanism to explain the present DM cosmological density.~Specifically, the relic abundance of SIMPs is set by so-called $3 \rightarrow 2$ processes and $2 \rightarrow 2$ forbidden annihilations, whereas in the simplified models we consider here the relic abundance is set by pair annihilation into visible particles.   

\subsection{Simplified models}
\label{sec:simp}
We start our exploration of spin-1 DM by considering a general set of simplified models~\cite{Catena_2018}. These models extend the Standard Model (SM) of particle physics by a complex vector field, $X^\mu$, playing the role of DM, and a mediator particle described by the real vector field $A'^\mu $. The following Lagrangian specifies the interactions between the complex and real vector fields $X^\mu$ and $A'^\mu$, respectively, and the Dirac spinors $f$ associated with the electrically charged SM fermions,
\begin{align}
    \mathscr{L} = &-\left[i b_5 X_\nu^\dagger \partial_\mu X^\nu A'^\mu + b_6 X_\mu^\dagger \partial^\mu X_\nu A'^\nu + h.c. \right] 
    \nonumber \\ 
    &-\left[b_7 \epsilon_{\mu \nu \rho \sigma} \left( X^{\dagger \mu} \partial^\nu X^\rho \right) A'^\sigma + h.c. \right] \nonumber \\ 
    &-h_3 A'_\mu \bar{f}\gamma^\mu f , \label{eq:Lsimplified}
\end{align}
where, as anticipated, $f$ includes electrons, muons, taus, and quarks (neutrinos are not included).~The first line in Eq.~(\ref{eq:Lsimplified}) describes interactions that can be generated in models for non-abelian spin-1 DM, as reviewed in Sec.~\ref{sec:non-abelian} and shown in detail in~\cite{Choi:2019zeb}.~The strength of these interactions is parametrized by the coupling constants $b_5$ and $b_6$.~Without loss of generality $b_5$ can be taken to be real, while the coupling constant $b_6$ is in general complex.~The second line in Eq.~(\ref{eq:Lsimplified}) describes interactions that can arise from abelian spin-1 DM models~\cite{Arcadi:2017kky}, and is characterised by the in general complex coupling constant $b_7$.~Finally, the last line in Eq.~(\ref{eq:Lsimplified}) corresponds to the coupling between the electrically charged SM fermions and a ``dark photon'', here associated with the vector field $A'^{\mu}$.~In order to make the analogy with the dark photon model explicit, one could identify $h_3$ with $h_3 = e\epsilon$, where,  $\varepsilon$ is the so-called kinetic mixing parameter, which enters the Lagrangian of the dark photon model via the term $-(\varepsilon/2) F_{\mu\nu} F'^{\mu \nu}$, $F_{\mu\nu}$ and $ F'^{\mu \nu}$ being the field strength tensors of the ordinary and dark photon, respectively.~In our numerical applications, we consider the following cases in which only one $b$-coupling at a time and $h_3$ are non-zero: $(h_3, b_5)\neq 0$, $(h_3,\Re[b_6])\neq 0$, $(h_3,\Im[b_6])\neq 0$, $(h_3,\Re[b_7])\neq 0$, and $(h_3,\Im[b_7])\neq 0$. To further facilitate the comparison between our results and the existing literature on the dark photon model, we later write the non-zero DM couplings in terms of $\alpha_D = g_D^2/(4\pi)$, where $g_D$ is one of $b_5$, $\Re[b_6]$, $\Im[b_6]$, $\Re[b_7]$ or $\Im[b_7]$. As far as the DM and mediator particle mass are concerned, we denote them by $m_X$ and $m_{A'}$, respectively.~In the simplified models of this section, they are free, independent parameters.

In this analysis, we restrict ourselves to the case where $m_{A'} \geq 2m_X$. Under this mass hierarchy, and since $g_D \gg e\epsilon$, dark photons are produced on shell and decay dominantly to DM at fixed target experiments. This choice allows for easy comparison with previous literature on invisible signatures at fixed target experiments \cite{LDMX,Berlin:2018bsc}. If $m_{A'} \leq 2m_X$, the predicted signal yield for missing energy/momentum signatures is suppressed, resulting in less sensitive predicted exclusion bounds \cite{Berlin_2019,PhysRevD.102.095011} and visible dark photon decay signatures becoming more relevant \cite{Berlin_2019}.

\subsection{Non-abelian SIMPs}
\label{sec:non-abelian}
As a second framework for vector DM, we consider a model where the DM candidate is a Strongly Interacting Massive Particle, or SIMP~\cite{Choi:2019zeb}.~In this model, the SM gauge group is extended by a local $SU_X(2)\times U(1)_{Z'}$ symmetry group under which none of the SM particles is charged.~The corresponding gauge couplings are $g_X$ and $g_{Z'}$.~The model also features an extended Higgs sector including a scalar singlet, $S$, and a second scalar $H_X$ transforming non-trivially under $SU_X(2)\times U(1)_{Z'}$.~The gauge bosons associated with the new symmetry group are denoted by $X_{i,\mu}$, $i=1,2,3$ and $Z'_\mu$, or, equivalently, by $X_\mu\equiv (X_{1,\mu}+ i X_{2,\mu})/\sqrt{2}$, $X^\dagger_\mu\equiv (X_{1,\mu} - i X_{2,\mu})/\sqrt{2}$, $X_{3,\mu}$ and $Z'_\mu$, respectively. The $SU(2)\times U(1)_{Z'}$ gauge group is spontaneously broken by the vacuum expectation values of $S$, i.e.~$v_S$, and $H_X$, i.e.~$v_X$.~This generates one complex and two real mass eigenstates, corresponding to $X_\mu$ and two linear combinations of $Z'_\mu$ and $X_{3,\mu}$, denoted here by $\tilde{Z}'_\mu$ and $\tilde{X}_{3,\mu}$, respectively. Their masses are~\cite{Choi:2019zeb},
\begin{align}
m^2_{X} &= \frac{1}{2} g_X^2 I v_X^2 \,, \nonumber\\
m^2_{ {\tilde Z}'} &=g^2_X I^2 v_X^2 \left(1-\cot\theta'_X\, g_{Z'}/g_X\right) \,, \nonumber\\
m^2_{{\tilde X}_{3}} &= g^2_X I^2 v_X^2 \left(1+\tan\theta'_X\, g_{Z'}/g_X\right) \,,
\label{eq:masses}
\end{align}
where 
\begin{align}
\tan(2\theta'_X)= \frac{2c_X s_X}{c^2_X-\alpha s^2_X}\,,
\end{align}
with $s_X=g_{Z'}/\sqrt{g_X^2+g_{Z'}^2}$, $c_X=g_{X}/\sqrt{g_X^2+g_{Z'}^2}$ and, finally $\alpha\equiv 1+q^2_S v^2_S/(I^2 v^2_X)$.~Here $q_S$ is the $U_{Z'}(1)$ charge of $S$, while $I$ labels the representation of $SU_X(2)$ under which $H_X$ transforms, e.g. $I=1/2$ for a doublet, $I=1$ for a triplet, and $I=3/2$ for a quadruplet.~The interaction Lagrangian containing the cubic self-interactions between these mass eigenstates is~\cite{Choi:2019zeb}, 
\begin{align}
\mathscr{L}_3 = &-i g_X\cos\theta'_X \bigg[
\left(\partial^\mu X^\nu -\partial^\nu X^\mu\right)
 X^\dagger_\mu {\tilde X}_{3,\nu} -
\left(\partial^\mu X^{\nu\dagger} -\partial^\nu X^{\mu\dagger}\right)
 X_\mu {\tilde X}_{3,\nu} \nonumber \\
& + X_\mu X^\dagger_\nu\left(\partial^\mu {\tilde X}^\nu_3 -\partial^\nu {\tilde X}^\mu_3\right)
\bigg] \nonumber \\
&-i g_X\sin\theta'_X \bigg[
\left(\partial^\mu X^\nu -\partial^\nu X^\mu\right)
 X^\dagger_\mu {\tilde Z}'_{\nu} -
\left(\partial^\mu X^{\nu\dagger} -\partial^\nu X^{\mu\dagger}\right)
 X_\mu {\tilde Z}'_{\nu} \nonumber \\
& + X_\mu X^\dagger_\nu\left(\partial^\mu {\tilde Z}^{\prime\nu} -\partial^\nu {\tilde Z}^{\prime\mu}\right)\bigg]. 
\label{eq:cubic}
\end{align}
The model also predicts quartic interactions between mass eigenstates.~These are relevant in relic density calculations, and have explicitly been calculated in~\cite{Choi:2019zeb}.~Finally, ``neutral current'' interactions between  $\tilde{Z}^\prime_\mu$, $ \tilde{X}_{3 \mu}$ and the charged SM fermions arise from a kinetic mixing term added to the Lagrangian of this $SU_X(2)\times U(1)_{Z'}$ gauge model.~Specifically, they are given by~\cite{Choi:2019zeb}
\begin{align}
\mathscr{L}_{\rm mix}&=  -e \varepsilon \cos (\theta_X') \, \tilde{Z}^\prime_\mu \, \bar{f}\gamma^\mu f  + e \varepsilon \sin (\theta_X') \, \tilde{X}_{3 \mu} \,\bar{f}\gamma^\mu f  \,.
\label{eq:Lmix}
\end{align}
Eq.~(\ref{eq:masses}) shows that for $\tan \theta'_X<0$ and $g_{Z'}/g_X |\tan \theta'_X|<1/2$, the predicted mass hierarchy is 
\begin{align}
m_X^2< m^2_{\tilde{X}_3} <m_{\tilde{Z}'}^2\,.
\label{eq:hierarchy}
\end{align}
For example, for $\sin (2 \theta_X') = -0.1$ and $\alpha_D\equiv g_X^2/(4 \pi)=0.5$, Eq.~(\ref{eq:hierarchy}) is always satisfied for perturbative values of $g_{Z'}$.~Consequently, when the mass hierarchy in Eq.~(\ref{eq:masses}) is realised, $X_\mu$ is a stable DM candidate, while $\tilde{X}_3$ and $\tilde{Z}'$ mediate the interactions between $X_\mu$ and the SM fermions via Eq.~(\ref{eq:Lmix}).~Finally, let us note that for $\cos (\theta'_X)=1$, Eq.~(\ref{eq:cubic}) reduces to the first line in Eq.~(\ref{eq:Lsimplified}) with $b_5=-2 \Im[b_6]=g_X$, if one integrates by part the second line in Eq.~(\ref{eq:cubic}) and uses the equation of motion for $X_\mu$ (i.e. Proca equation), which implies $\partial_\mu X^\mu=0$.

\subsection{Vector DM phenomenology: general considerations}
In this subsection, we highlight the main differences between the models defined in Sec.~\ref{sec:simp} and in Sec.~\ref{sec:non-abelian} focusing on the predicted DM relic abundance, kinetic equilibrium, mediator production and decay rates, and, finally, relativistic DM-electron and -nucleon scattering cross sections.

\subsubsection{Relic density}
Let us denote the DM particle associated with $X_\mu$ by $X^{+}$ and the corresponding DM antiparticle by $X^{-}$.~Also, let $n_{X^{+}}$ be the cosmological number density of DM particles and $n_{X^-}$ be the corresponding cosmological density of DM antiparticles.~In the case of the CP-preserving simplified or SIMP DM models that we consider here, $n_{X^{+}}=n_{X^{-}}$, and the total DM number density, $n_X=n_{X^{+}}+n_{X^{-}}$, evolves with time according to the following Boltzmann equation \cite{Choi:2019zeb},
\begin{align}
\dot{n}_X + 3H n_X = &-\frac{1}{2}\langle \sigma v_{\rm rel} \rangle_{X^+X^-\rightarrow f{\bar f}}\, (n_X^2 - n_{X,{\rm eq}}^2)  
\nonumber \\
&- \frac{1}{2}\langle \sigma v^2_{\rm rel} \rangle_{X^+X^+X^-\rightarrow X^+\tilde{X}_3}\, \left(n_X^3 - n_Xn_{X,{\rm eq}}^2\right) \nonumber \\
&+ 2\,\langle \sigma v_{\rm rel} \rangle_{\tilde{X}_3\tilde{X}_3\rightarrow X^+X^-} \, n_{\tilde{X}_3,{\rm eq}}^2\left(
1-\frac{n^2_X}{n^2_{X,{\rm eq}}} \right) \,.
\label{eq:Boltz}
\end{align}
In general, Eq.~(\ref{eq:Boltz}) shows that $n_X$ evolves from an equilibrium configuration to a constant co-moving value as a result of the DM chemical decoupling from the thermal bath, i.e.~the so-called freeze-out or chemical decoupling mechanism.~The first term in the right-hand-side of Eq.~(\ref{eq:Boltz}) describes the time evolution of the DM number density due to DM pair annihilation into SM fermions.~For $\epsilon\sim 10^{-4}$ ($10^{-6}$) and $m_X\sim~100$~MeV ($1$~MeV), this term alone can account for the entire DM relic density.~We will explore this DM production channel within the framework of simplified models for vector DM introduced above.~The second and third line in Eq.~(\ref{eq:Boltz}) describes the time evolution of the DM number density due to $3 \rightarrow 2$ processes and $2 \rightarrow 2$ forbidden annihilations that are specific to the vector SIMP model introduced above (i.e.~they are zero in the simplified model framework).~The DM to $\tilde{X}_3$ mediator mass ratio determines whether the DM relic abundance is set by $3 \rightarrow 2$ processes or $2 \rightarrow 2$  forbidden annihilations, as shown in Fig.~3 of \cite{Choi:2019zeb}.~We will explore the interplay of these two DM production mechanisms in the case of SIMP DM. 

\subsubsection{Kinetic equilibrium}
In Eq.~(\ref{eq:Boltz}) we implicitly assumed the mass hierarchy of Eq.~(\ref{eq:hierarchy}) for the SIMP DM model we introduced in Sec.~\ref{sec:non-abelian}.~We also assumed that the associated mediator ${\tilde X}_3$ is in kinetic equilibrium with the SM thermal bath during the DM freeze-out.~Interestingly, the in-equilibrium decay of ${\tilde X}_3$ into SM particles, combined with the effective scattering of DM with the ${\tilde X}_3$ mediator induced by the cubic and quartic interactions introduced above, serve as mechanisms to keep the DM particles in kinetic equilibrium during freeze-out, and thereby satisfy the strong constraints from structure formation on the DM kinetic decoupling temperature~\cite{Choi:2019zeb}.~For a given $g_X$, $\sin(2 \theta'_X)$, $m_X$ and $m_{\tilde{X}_3}$, the value of $\epsilon$ required for ${\tilde X}_3$ and, consequently, for the DM particles to be in kinetic equilibrium at the freeze-out temperature $T_f$ can be estimated from 
\begin{align}
n_{\tilde{X}_3,\rm eq}(T_f) \Gamma_{\tilde{X}_3} > H(T_f)n_{X,\rm eq}(T_f)\,,
\label{eq:kin}
\end{align}
where $H(T_f)$ is the Hubble rate at $T_f$, and the equilibrium densities for $X$ and ${\tilde X}_3$ are given by
\begin{align}
n_{X,{\rm eq}} &= \frac{45x^2}{2g_{*s}(T)\pi^4}\, s\,K_2(x)\,, \nonumber\\
n_{{\tilde X}_3,{\rm eq}} &=  \frac{45 m_{\tilde{X}_3}^2x^2}{4g_{*s}(T)\pi^4 m_X^2}\, s\,K_2\Big(\frac{m_{\tilde{X}_3}x}{m_X}\Big) \,,
\end{align}
where $s$ is the entropy density, $g_{*s}(T)$ is the effective number of entropic relativistic degrees of freedom at the temperature $T$, $K_2$ is a modified Bessel function of the second kind and $x\equiv m_X/T$.~In Eq.~(\ref{eq:kin}), $\Gamma_{\tilde{X}_3}$ is the total decay rate of $\tilde{X}_3$, for which the expression is given in~\cite{Choi:2019zeb}. \\

\subsubsection{Mediator production and decay}
Key to our exploration of spin-1 DM is the production and subsequent decay of mediator particles in fixed target and beam dump experiments.~In the case of the simplified models of Sec.~\ref{sec:simp}, we generically expect that $A'$ particles are produced through the kinetic mixing term in Eq.~(\ref{eq:Lsimplified}), either via dark bremsstrahlung or via meson decays.~These $A'$s are then expected to decay into a DM particle/anti-particle pair for values of $\epsilon$ that are consistent with the observed relic density set via the first term in the right-hand side of Eq.~(\ref{eq:Boltz}).~In contrast, in the case of the vector SIMP DM model introduced in Sec.~\ref{sec:non-abelian}, both the $\tilde{X}_3$ and the $\tilde{Z}'$ mediator can in principle be produced in fixed target and beam dump experiments.~In our calculations we will focus on $\sin(2\theta'_X)=-0.1$, which implies that only $\tilde{Z}'$ particles can significantly be produced via the interaction in Eq.~(\ref{eq:Lmix}), and that, once produced, these $\tilde{Z}'$ particles will dominantly decay invisibly via gauge bosons self-interactions, as long as $\epsilon$ is small compared to $g_X$.~Consequently, for $\sin(2\theta'_X)=-0.1$, the simplified and SIMP models behave similarly from the point of view of ``dark vector boson'' production at fixed target and beam dump experiments.

\subsubsection{Scattering by electrons and nuclei}
Finally, we also need the (relativistic) cross sections for DM-electron and -nucleon scattering in order to compare the simplified models of Sec.~\ref{sec:simp} and the SIMP DM model of Sec.~\ref{sec:non-abelian} with the experimental constraints reviewed in Sec.~\ref{sec:bounds}.~This in particular applies to the analysis of DM direct detection experiments and of beam dump experiments, where DM particles produced by the decay of $A'$ mediators (in the case of simplified models) or $\tilde{Z}'$ mediators (in the case of SIMPs)  are searched for in electron or nuclear recoil events in a downstream detector.

We calculate these cross sections by implementing the models of Sec.~\ref{sec:theory} in {\sffamily FeynRules}~\cite{Alloul:2013bka} and then using {\sffamily CalcHEP}~\cite{Belyaev:2012qa} to generate analytic expressions for the squared modulus of the spin-averaged scattering amplitudes.~We finally validate the outcome of this symbolic calculation through direct analytical calculations of a subset of selected cross sections.~In appendices~\ref{sec:sigmasimp} and~\ref{sec:sigmasimpN}, we list the relativistic DM-electron and -nucleon scattering cross sections that we find for the simplified models of Sec.~\ref{sec:simp}.~In appendix~\ref{app:non-abelian}, we report the scattering cross sections that we obtain for the SIMP DM model of Sec.~\ref{sec:non-abelian} as explained above.~In the calculation of DM-nucleon scattering cross sections, we use the mediator-nucleon interaction Lagrangian,
\begin{align}
\mathscr{L}_{N} =e\epsilon F_1\Bar{\psi} \gamma^\mu \psi A'_\mu  + e\epsilon \frac{F_2}{2 m_N} \left[ \Bar{\psi} \sigma^{\mu \nu} \left(\partial_\nu  \psi \right) 
 +  \left (\partial_\nu \Bar{\psi} \right) \sigma^{\mu \nu} \psi \right]A'_\mu \,,
 \label{eq:LN}
\end{align}
in the case of simplified models. For the case of SIMP DM, we use Eq.~(\ref{eq:LN}) but with $F_1 (F_2) \rightarrow \cos(\theta'_X) F_1 (F_2)$ and $A'\rightarrow \tilde{Z'}$.~Here, $F_1$ and $F_2$ are nuclear form factors, and we list them in appendix~\ref{sec:sigmasimpN}.

\section{Experimental and theoretical constraints}
\label{sec:bounds}
In this section, we introduce a selection of constraints and projections from, respectively, operating and future DM search experiments that apply to the models of Sec.~\ref{sec:theory}.~By complementing these constraints and projections with bounds from the unitarity of the S-matrix, in Sec.~\ref{sec:results} we will identify the regions of the parameters space of the vector DM models we consider in this work where DM is simultaneously:~1) thermally produced, 2) experimentally allowed and, finally 3) detectable.~We refer to these regions in parameter space as  {\it thermal targets}.

\begin{figure}[t]
    \centering
    \includegraphics[width=0.8\linewidth]{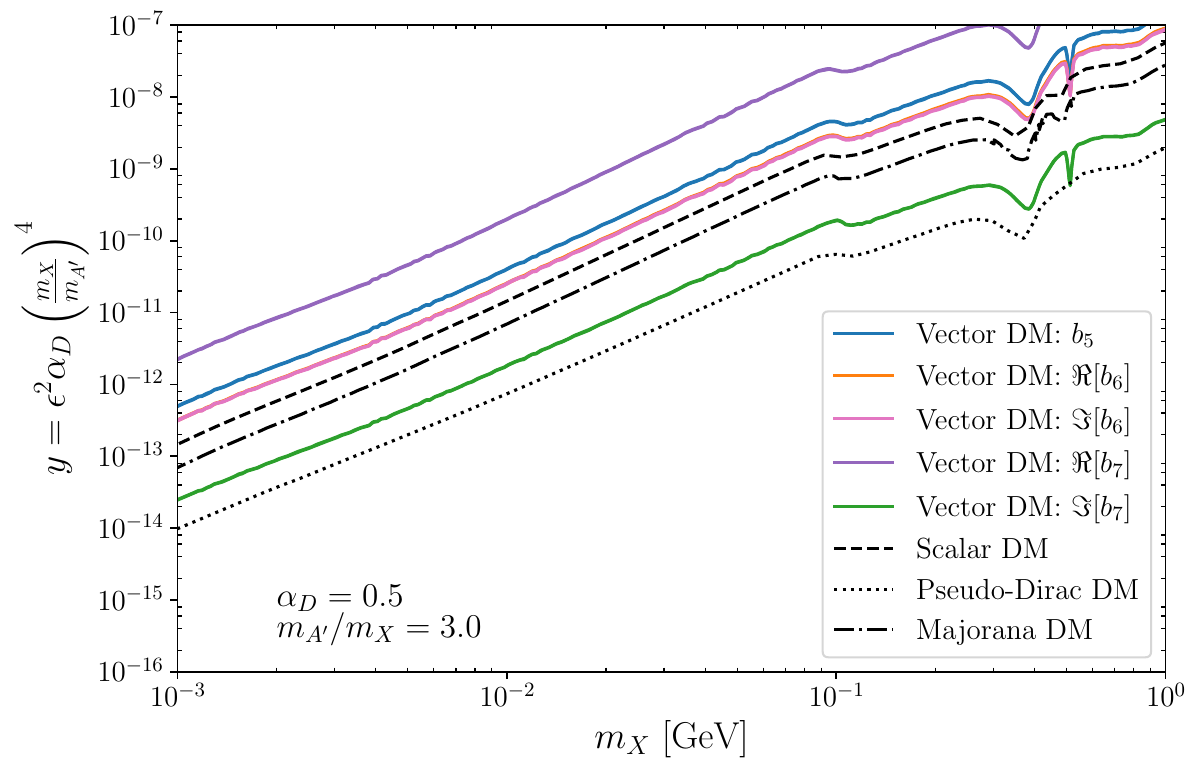}
    \caption{Contours consistent with the observed DM abundance for each of the simplified models for spin-1 DM introduced in Sec.~\ref{sec:simp} (coloured solid contours) and for scalar, pseudo-Dirac, and Majorana DM (dashed, dotted, and dash-dot) with $m_{A'}=3m_X$, and $\alpha_D = g_D^2/(4 \pi) = 0.5$, where $g_X$ is the corresponding $b$-coupling in the legends.~Results are presented in the plane spanned by the DM particle mass and the standard ``coupling constant'' $y$.}
    \label{fig:RelicContours}
\end{figure}

\subsection{Relic density}
Accurate measurements of the Cosmic Microwave Background (CMB) angular power spectrum by the Planck collaboration set strong constraints on the present DM cosmological density~\cite{Planck:2018vyg}.~The spin-1 DM models considered in this work, introduced in Sec.~\ref{sec:theory}, are capable of producing the observed DM relic abundance consistent with Planck by the freeze-out mechanism.~In the case of simplified models (Sec.~\ref{sec:simp}), the freeze-out of DM pair annihilations into visible SM particles 
sets the DM relic density, whereas in the case of vector SIMP DM (Sec.~\ref{sec:non-abelian}) the present DM cosmological density arises from the freeze-out of $3\rightarrow 2$ and forbidden annihilations, as one can see from Eq.~(\ref{eq:Boltz}).~Below, we discuss the two scenarios separately.

In the case of simplified models, we are interested in the region of parameter space where $m_{A'}>2m_X$. In this region, the relic abundance is set dominantly by direct DM annihilation into SM fermions through an s-channel. Comparing our theoretical predictions based on Eq.~(\ref{eq:Lsimplified}) with CMB data, we calculate the thermally averaged cross sections for direct annihilation using the {\sffamily MicrOMEGAS} software \cite{Belanger:2006is}. We then compute the DM relic density by using our own Boltzmann solver, which relies on the freeze-out approximation from \cite{Gondolo:1990dk}. Our software results agree with {\sffamily MicrOMEGAS}, although we include the contributions from DM annihilation into hadronic final states~\cite{Izaguirre_2016,PDG,serendipity}.~Fig. \ref{fig:RelicContours} shows the contour lines consistent with the Planck DM abundance, or relic targets, for each of the simplified models introduced in Sec.~\ref{sec:simp} in addition to three  benchmark models from \cite{Berlin_2019} (and mentioned in the introduction, namely complex scalar, pseudo-Dirac, and Mjorana DM). Each of the simplified models for spin-1 DM in Fig.~\ref{fig:RelicContours} is defined as having all couplings in Eq.~(\ref{eq:Lsimplified}) set to 0 except for $h_3$ and the $b$-coupling specified in the legends. There are kinks in the curves occurring at $m_X \approx m_{\mu}$, where DM annihilation into muons become kinematically accessible. The resonance features appear because of resonances in the cross sections for DM annihilation into hadrons.

\begin{figure}[t]
    \centering
    \includegraphics[width=0.8\linewidth]{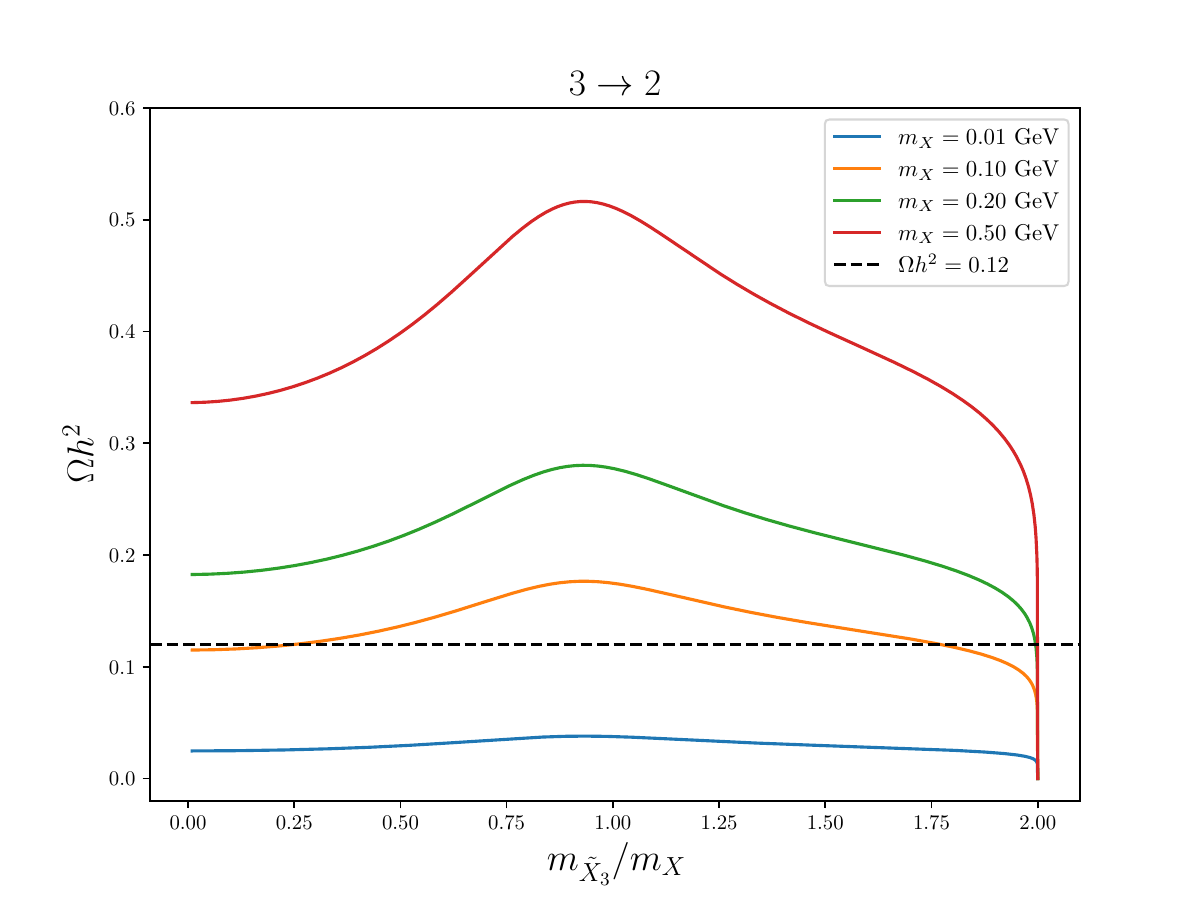}
    \caption{DM relic abundance, $\Omega h^2$, as a function of $m_X$ and $m_{\tilde{X}_3}/m_X$ for SIMP DM thermally produced by the freeze-out of $3\rightarrow 2$ processes.~Here we take $\alpha_D = 0.5$.~The horizontal dashed line corresponds to the observed value of $\Omega h^2$.~Coloured lines have been obtained by evaluating Eq.~(40) from~\cite{Choi:2019zeb}.}
    \label{fig:relicSIMP}
\end{figure}
In the case of vector SIMP DM, we rely on previus results from~\cite{Choi:2019zeb}.~For each given $\alpha_D=g_X^2/(4\pi)$ ($g_X$ is one of the gauge couplings introduced in Sec.~(\ref{sec:non-abelian})) and $\sin(2\theta'_X)$ we extract the value of the $m_{\tilde{X}_3}/m_X$ ratio that gives the correct DM relic density from Fig.~3 in~\cite{Choi:2019zeb}.~While for $\alpha_D=0.1$ the correct relic abundance can be obtained for DM masses above approximately 10 MeV, for $\alpha_D=0.5$ vector SIMP DM can account for the whole cosmological abundance of DM only for masses above about 80 MeV.~This effect is illustrated in Fig.~\ref{fig:relicSIMP}, which shows the DM relic abundance as a function of the DM mass and of the $m_{\tilde{X}_3}/m_X$ ratio, focusing on the case in which $3\rightarrow 2$ are the dominant production mechanism.~As one can see from this figure, for $m_{\tilde{X}_3}/m_X=2$ the DM relic abundance drops to zero, independently of the DM particle mass.~This is due to the fact that for $ m_{\tilde{X}_3}/m_X>2$, the $3\rightarrow 2$ process is kinematically not allowed. 

\subsection{Direct Detection}
In principle, DM direct searches via electronic transitions in detector materials located deep underground also place important constraints on sub-GeV DM models, e.g.~\cite{Catena_2023}. For spin-1 DM, the current most competitive constraints are from Xenon1T and Xenon10, which only appear in our plots for the $b_5$ and $\Re[b_7]$ models. We take the 90\% C.L. exclusion limits from the work of \cite{Catena_2023}.

\subsection{Energy Injection}
DM annihilations in the early universe can inject energetic particles into the photon-baryon plasma or into the intergalactic medium (IGM), altering the CMB or the IGM temperature, respectively. However, CMB limits are only competitive for s-wave annihilating DM \cite{CMB_swave,MeVDMComplementarity}. Among the models introduced in Sec.~\ref{sec:theory}, only the model with $\Im[b_7]\neq 0$ gives rise to s-wave dominant annihilation cross sections, thus we include 95\% C.L. exclusion bounds from the CMB for this model. For all remaining spin-1 models in Sec.~\ref{sec:theory}, including the SIMP DM model, the predicted annihilation cross section is p-wave.~CMB limits on p-wave annihilation cross sections have been computed here~\cite{Liu_2016}.~We find that they are weaker than the IGM limits on the spin-1 models we consider in this work.~Consequently, we set 95\% C.L. exclusion limits from measurements of the IGM temperature extracted from Lyman-$\alpha$ observations on our p-wave annihilating DM models.~Specifically, we require that the predicted thermally averaged annihilation cross section is smaller than the upper bounds reported in~\cite{Liu_2021} as a function of the DM mass.~In this analysis, we calculate the relevant annihilation cross sections analytically, and then compare them with the output of {\sffamily MicrOMEGAS}.~These IGM upper limits only appear in the top left corner of our parameter space for certain mass ratios ($m_{A'}/m_X$) and models.

\subsection{Beam Dumps and Fixed Target Experiments}
\label{sec:bd}
 A beam of protons or electrons incident on a fixed target creates cascades of interactions at beam dump experiments. The goal is to produce DM in these cascades, which can then be detected in a downstream detector. Calculating the experimental reach of these experiments involves modeling the processes that give rise to DM production and detection, which is done through Monte Carlo (MC) simulations. These simulations include the details of the detector, the beam, the interactions which produce DM, and the DM model. With this objective, we use a modified version of the numerical tool BdNMC, a beam dump Monte Carlo software package \cite{deNiverville_2017}, in which we implement our spin-1 DM models to simulate DM production and interactions. BdNMC has the benefit of providing a simple and rigorous framework for simulating the relevant experiments Mini-Boone and LSND. In the following section, 90\% C.L. limits on the model parameter space calculated from these simulations are presented, showing the reach of current experiments including beam dump and missing energy/momentum experiments which give rise to the most competitive constraints on spin-1 DM. Below, each experiment considered is introduced.

\subsubsection*{LSND and MiniBooNE}

At proton beam dump experiments LSND \cite{deNiverville:2011it,LSND:2001akn} and MiniBooNE \cite{MiniBooNEDM:2018cxm}, a proton beam is incident on a target and the following chain of interactions occur producing DM: $p p \to X \pi^0 ; \pi^0 \to \gamma A'; A' \to DM DM$, where in the inclusive process, $p p \to X \pi^0$, $X$ denotes an unspecified/unmeasured set of particles, whereas $\pi^0$ can also be $\eta$ in the case of MiniBooNE where the energy is sufficient. In addition, dark proton bremsstrahlung also produces DM in the case of MiniBooNE. The produced DM can then be detected in the downstream detector through DM-electron and for MiniBooNE also DM-nucleon scattering.
We compute the expected number of signal events using a modified version of the software BdNMC \cite{deNiverville:2016rqh}, a MC simulation tool for beam dumps. We added to BdNMC the model dependent DM-electron and -nucleon scattering cross section reported in the App.~\ref{sec:sigmasimpN} and \ref{sec:sigmasimp} as well as the relevant branching ratios for dark photon decay.

At LSND, 55 non standard events were observed at 90\% confidence level. A factor of 2 is included to account for the uncertainty in the pion production rate, so we take the 90\% C.L. at 110 events. We perform MC simulations of DM events at LSND, taking into account our model dependent cross sections, and we draw the contours in $y$ vs $m_{X}$ space that corresponds to 110 signal events.~In this work, we adopt the standard notation $y=\epsilon^2 \alpha_D (m_X/m_{A'})^4$.

Similarly, we perform MC simulations of MiniBooNE to calculate the expected number of signal events. No events were observed at MiniBooNE, thus we take the contour in $y$ vs $m_{X}$ space at 2.3 events which corresponds to a 90\% C.L. exclusion limit.

\subsubsection*{E137}
Limits from the electron beam dump experiment E137 are not competitive with MiniBooNE and LSND, thus we do not include them in our analysis.

\subsubsection*{NA64}
The missing energy experiment NA64 \cite{NA64:2019imj}, where an electron beam is incident on a target, aims to produce a DM flux from dark bremsstrahlung and detect these signals by their missing energy. No signal events have been observed at this experiment. We report 90\% confidence limits extracted from  \cite{NA64:2019imj}, which we project onto the $y$ vs $m_X$ plane.

\subsubsection*{LDMX}
LDMX is a future missing momentum experiment~\cite{LDMX,Berlin:2018bsc} with an 8 GeV electron beam incident on a tungsten target with $10^{16}$ EOT. The ultimate reach at  90\% C.L. is calculated using the expected number of DM signal events from simulations and we project this expected limit onto our parameter space. In addition, we include the projected 90 \% C.L. exclusion limits from the analysis of \cite{Schuster:2021mlr}, where bremsstrahlung photons are converted into vector mesons and then decay to invisible states, giving an extended sensitivity reach in the larger DM mass regime.

\subsection{Monophoton searches}
We also include searches for single photon events at $e^+ e^-$ colliders, where DM is produced through the process $e^+ e^- \to \gamma A', A' \to DM DM$.

\subsubsection*{BaBar}
The BaBar detector at the PEP-II B-factory searches for a narrow peak in the missing mass distribution in the events with one high energy photon. BaBar has observed no signal events, and we take the 90\% C.L. limits on $\epsilon$ vs $m_X$ from  \cite{Lees:2017lec} and project them onto our parameter space.

\subsubsection*{Belle-II}
For the future experiment Belle-II, we use the expected 90\% C.L. limits on $\epsilon$ vs $m_X$ for phase 3 of Belle-II reported in \cite{belle2}, to obtain the expected exclusion limits on our spin-1 DM model.

\subsection{Unitarity bound}
In general, simplified DM models can violate perturbative unitarity in some regions of parameter space \cite{Kahlhoefer:2015bea}.~Due to the energy dependence introduced in the cross section by the longitudinal component of the vector DM polarization vectors, as well as from the underlying derivative couplings, the simplified models for spin-1 DM introduced in Sec.~\ref{sec:simp} predict large and un-physical scattering amplitudes. Therefore, their parameter space is subject to a ``unitarity bound'', or in other words a bound of theoretical validity.~Violations of this bound indicate either that the theory is non perturbative (all terms in the perturbative expansion of the S-matrix are equally important), or that it is not complete,
and thus additional fields have to be included to cancel out energy dependent, un-physical contributions to scattering cross sections. Unitarity violation from DM self-scattering was investigated in \cite{Chang:2023cki}, while here we calculate the DM-$e^-$ scattering amplitude to determine at which parameters unitarity is violated.~Following~\cite{Kahlhoefer:2015bea}, we implement the unitary bound by requiring that in the $(y,m_X)$ plane, 
\begin{align}
    \lvert \Im(M_{i\rightarrow i}^J) \rvert   \leq 1 \,,\nonumber\\
    2\lvert \Re(M_{i\rightarrow i}^J) \rvert \leq 1 \,,
\end{align}
where $i\rightarrow i = X e^- \to X e^- $, $J=0$, and
\begin{equation}
    \mathcal{M}_{X e^- \to X e^-}^0 (s) = \frac{\beta}{32 \pi} \int^1_{-1} d\cos\theta  \mathcal{M}_{X e^- \to X e^-}(s,\cos\theta) \,.
\end{equation}

\begin{figure}[t]
    \centering
    \begin{subfigure}[b]{0.49\textwidth}
    \includegraphics[width=\linewidth]{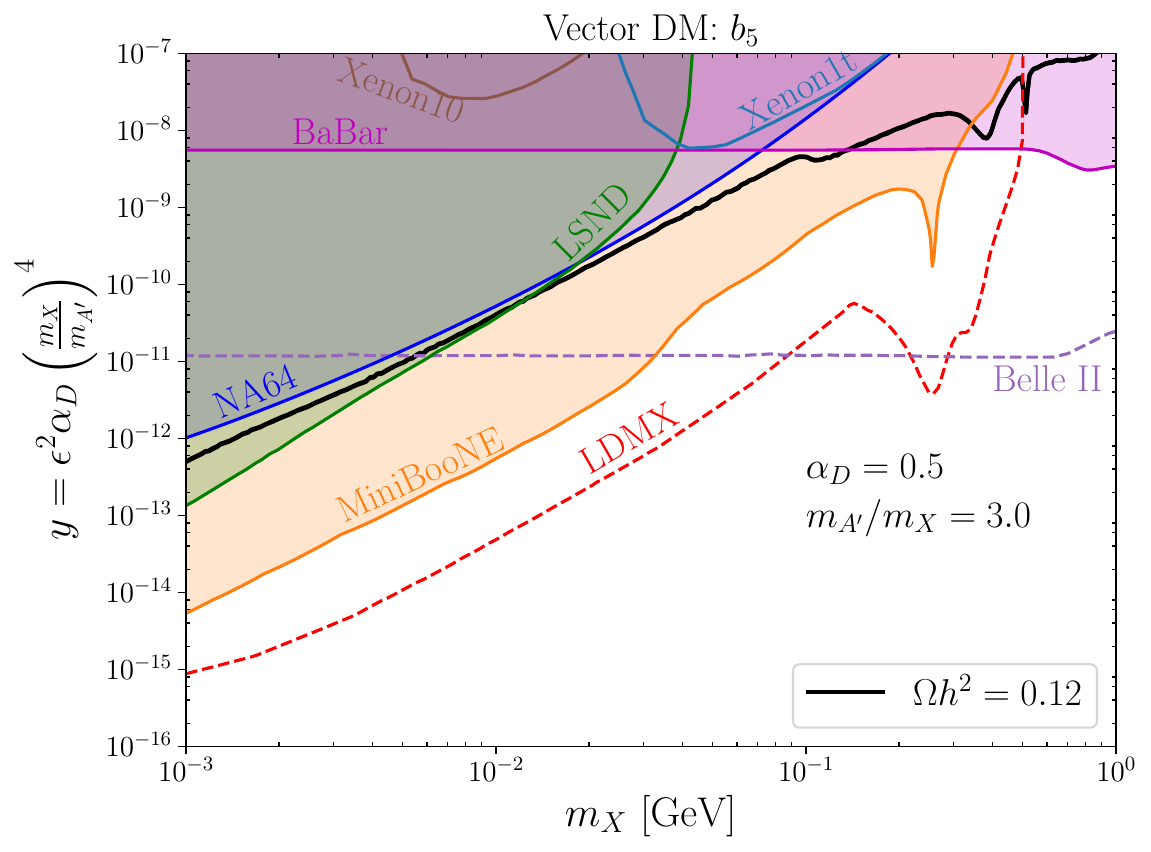}
    \caption{}
    \label{subfig:b5_ratio3}
    \end{subfigure}
    \begin{subfigure}[b]{0.49\textwidth}
    \includegraphics[width=\linewidth]{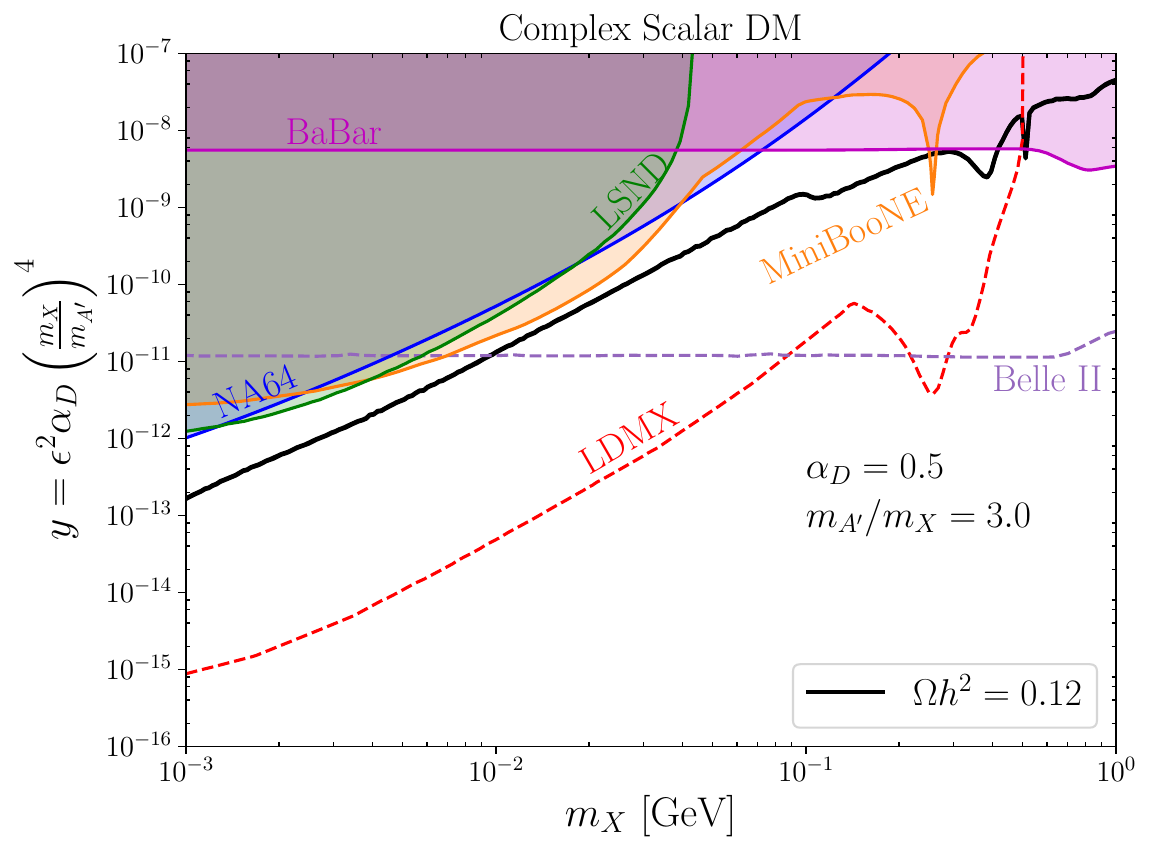}
    \caption{}
    \label{subfig:scalar_ratio3}
    \end{subfigure}
    \begin{subfigure}[b]{0.49\textwidth}
    \includegraphics[width=\linewidth]{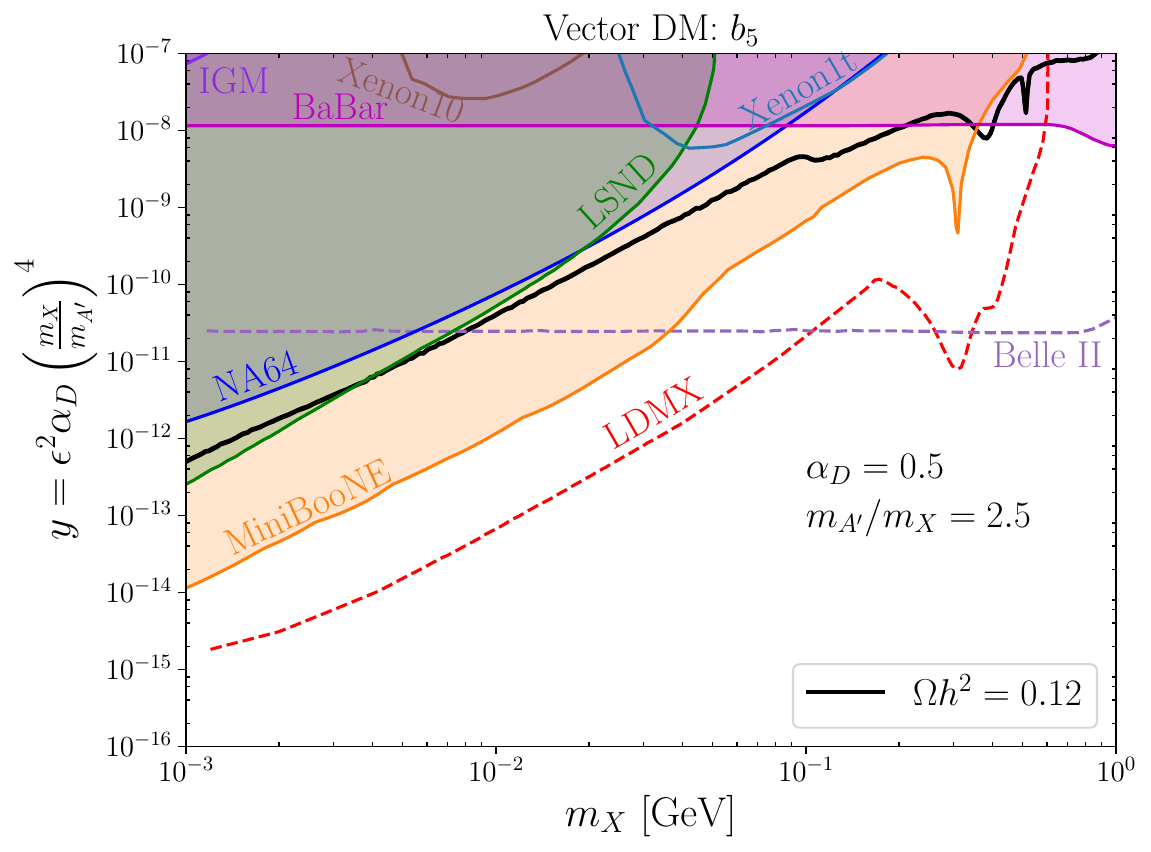}
    \caption{}
    \label{subfig:b5_ratio2.5}
    \end{subfigure}
    \begin{subfigure}[b]{0.49\textwidth}
    \includegraphics[width=\linewidth]{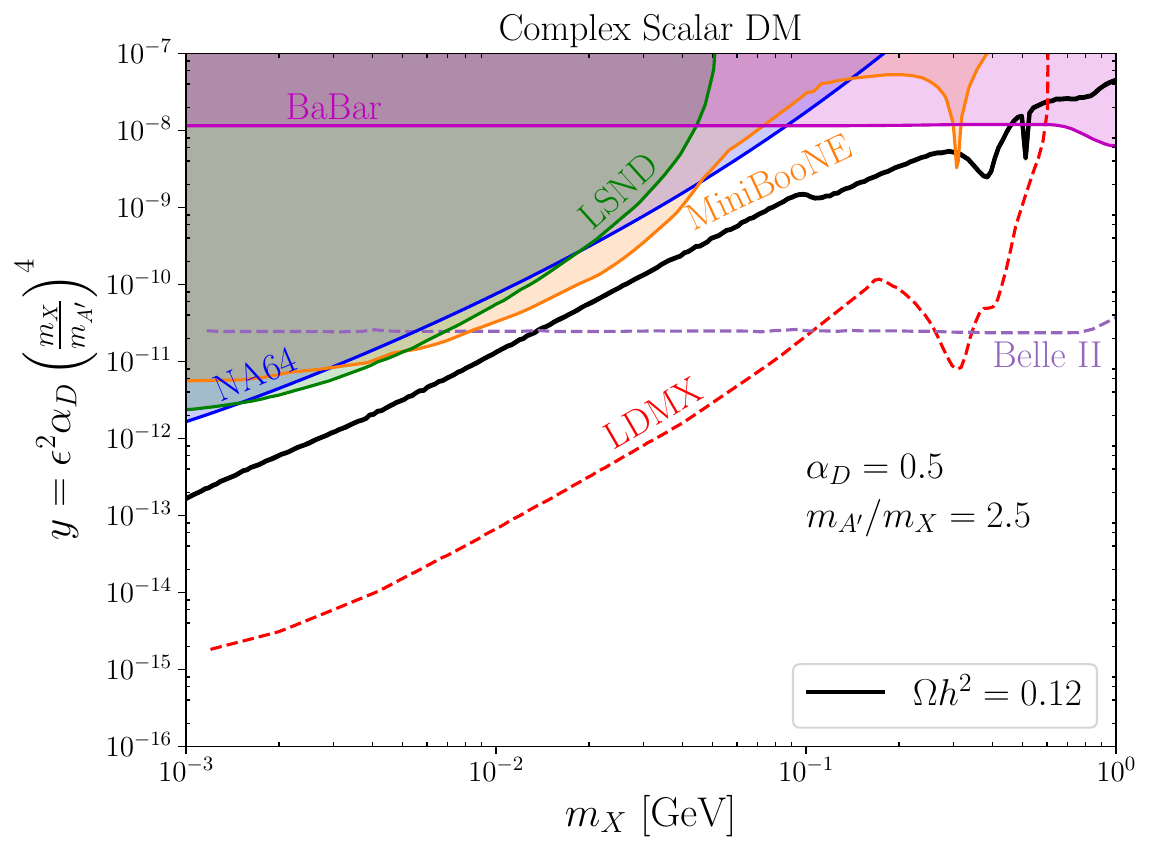}
    \caption{}
    \label{subfig:scalar_ratio2.5}
    \end{subfigure}
    \caption{{\it Panels (\ref{subfig:b5_ratio3}) and (\ref{subfig:b5_ratio2.5}}).~Exclusion limits from MiniBooNE, LSND, NA64, BaBar, IGM temperature observations and direct detection experiments (coloured shaded regions) and projected reach for Belle II and LDMX (dashed line) on the simplified model with $b_5\neq 0$ (see Sec.~\ref{sec:simp}) in the parameter space, $y$ vs $m_X$ for $\alpha_D=0.5$.~The contours consistent with the observed relic abundance ($\Omega h^2 \approx 0.12$) are drawn in solid black.~The top panel is for $m_{A'}=3m_X$, whereas the bottom panel assumes $m_{A'}=2.5m_X$.~The model is strongly constrained by MiniBooNE.~{\it Panels (\ref{subfig:scalar_ratio3}) and (\ref{subfig:scalar_ratio2.5}}).~Same as the left panels, now for complex scalar DM.
    }
    \label{fig:scalar_b5_limits}
\end{figure}

\begin{figure}[t]
   \centering
    \begin{subfigure}[b]{0.49\textwidth}
    \includegraphics[width=\linewidth]{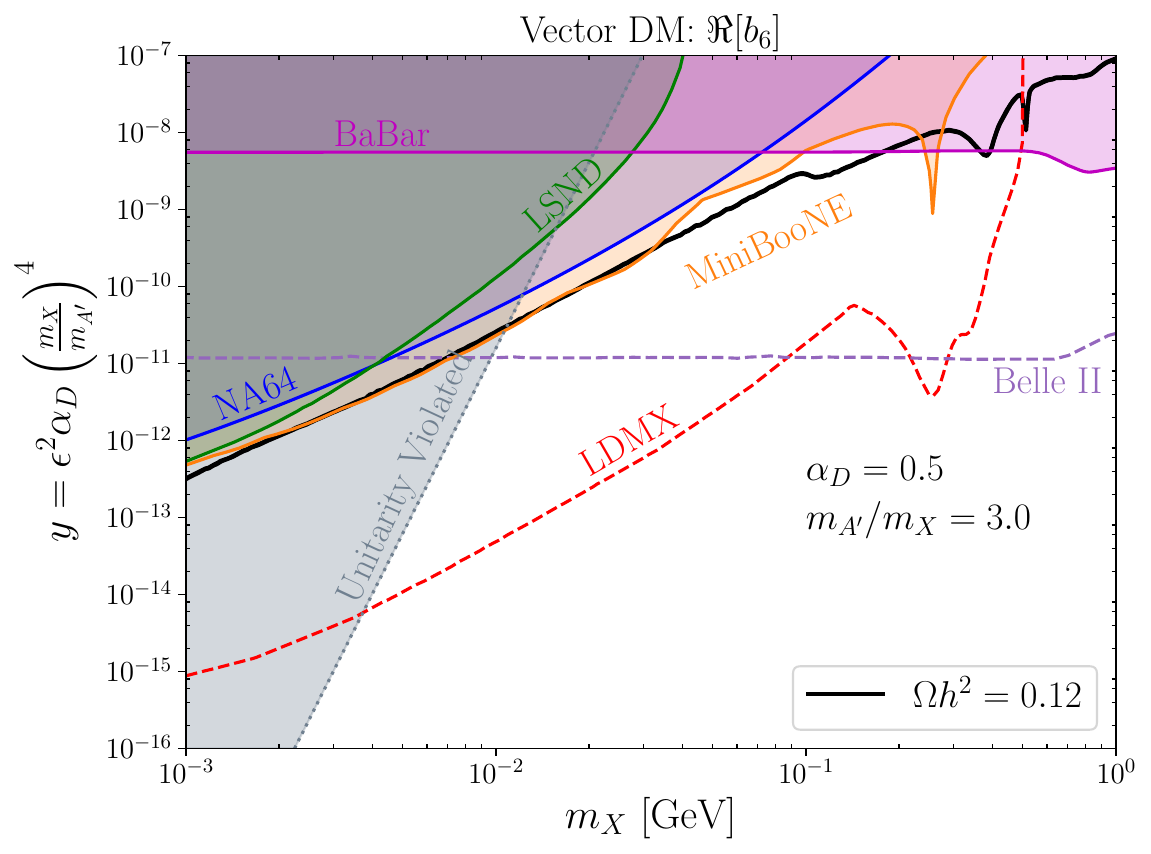}
     \caption{}
     \label{subfig:b6R_ratio3}
    \end{subfigure}
    \begin{subfigure}[b]{0.49\textwidth}
    \includegraphics[width=\linewidth]{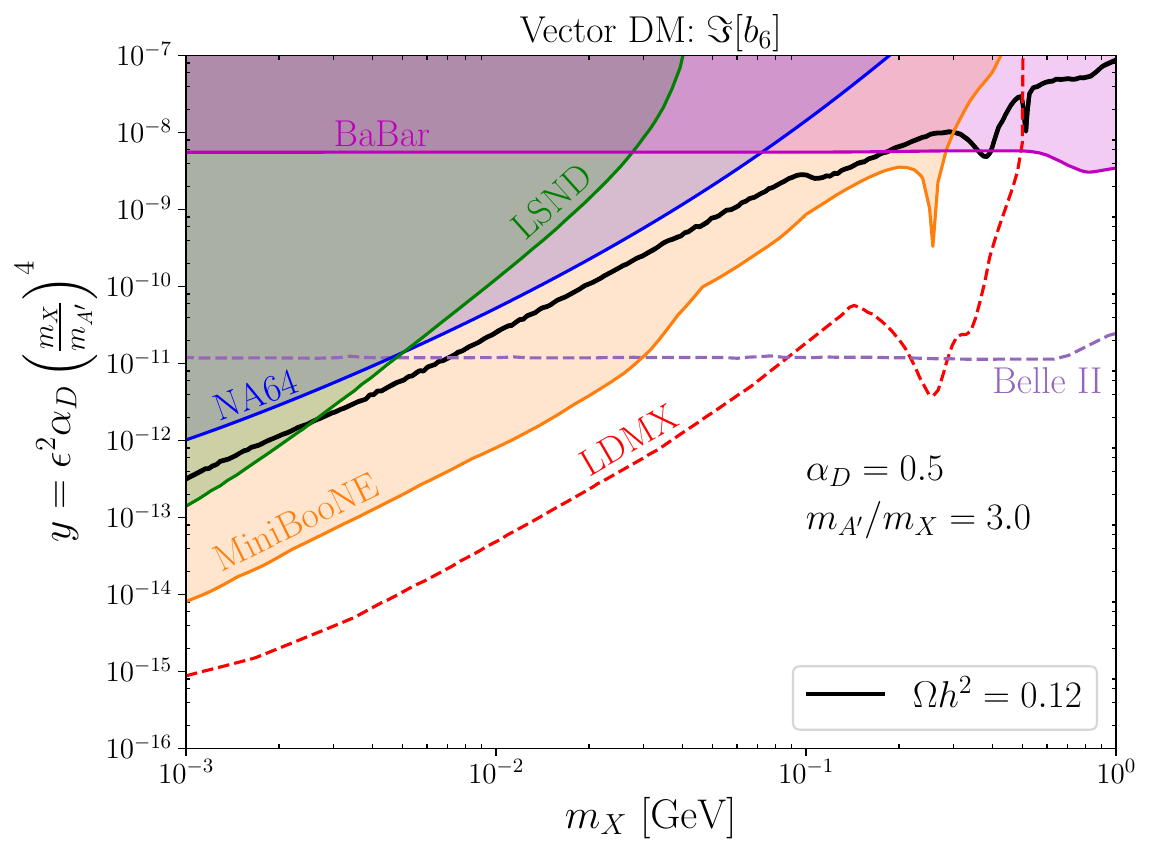}
     \caption{}
     \label{subfig:b6I_ratio3}
    \end{subfigure}
    \begin{subfigure}[b]{0.49\textwidth}
    \includegraphics[width=\linewidth]{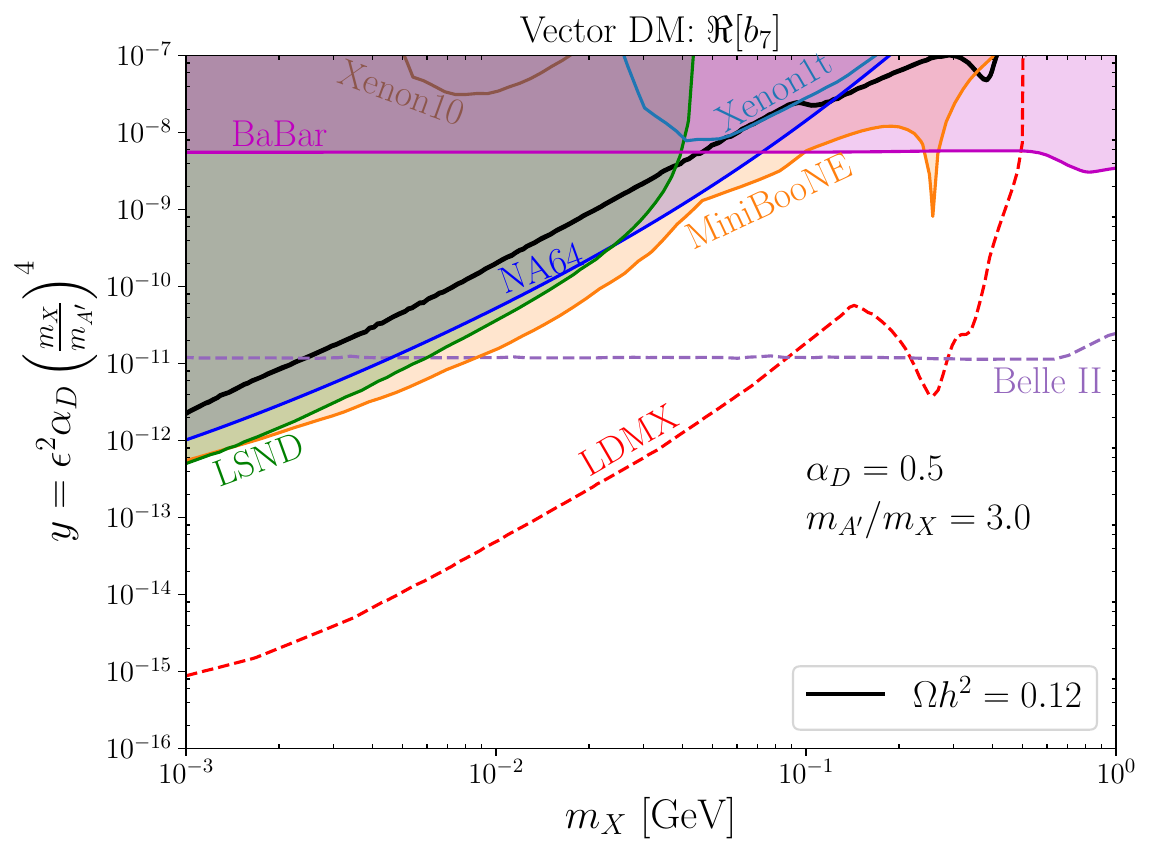}
     \caption{}
     \label{subfig:b7R_ratio3}
    \end{subfigure}
    \begin{subfigure}[b]{0.49\textwidth}
    \includegraphics[width=\linewidth]{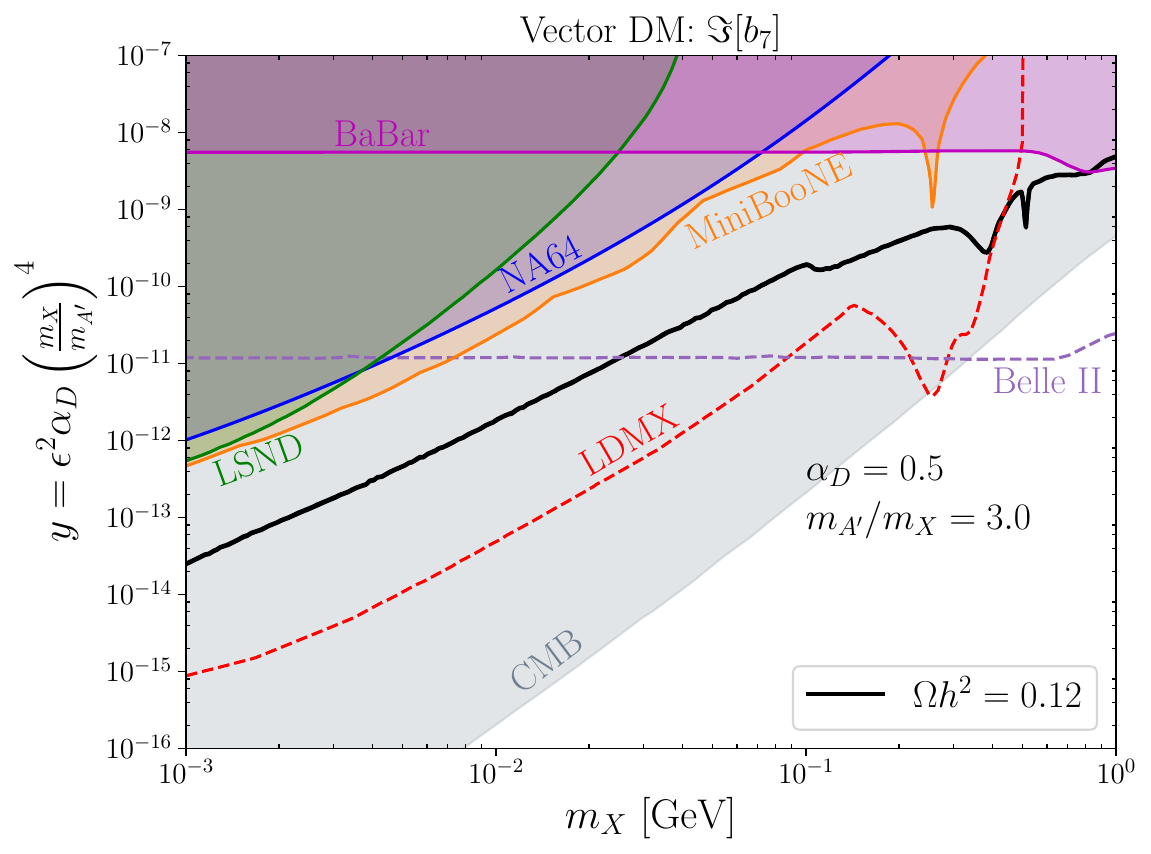}
     \caption{}
     \label{subfig:b7I_ratio3}
    \end{subfigure}
    \caption{Same as Fig.~\ref{fig:scalar_b5_limits}, now for the $\Re[{b_6}]$, $\Im[{b_6}]$, $\Re[{b_7}]$, and $\Im[{b_7}]$ models with $\alpha_D=0.5$ and $m_{A'}=3m_{\chi}$.~The models with $\Re[{b_7}] \neq 0$ and $\Im[b_6]\neq0$ are strongly constrained by beam dump experiments, while the model with  $\Im[{b_7}] \neq 0$ is ruled out by CMB observations.~The $\Re[b_6]$ model predicts the current DM relic abundance while also being compatible with observations and within reach at LDMX in the mass range between about 40 MeV and 200 MeV.
    }
    \label{fig:limits_mratio3}
\end{figure}

\begin{figure}[t]
   \centering
    \begin{subfigure}[b]{0.49\textwidth}
    \includegraphics[width=\linewidth]{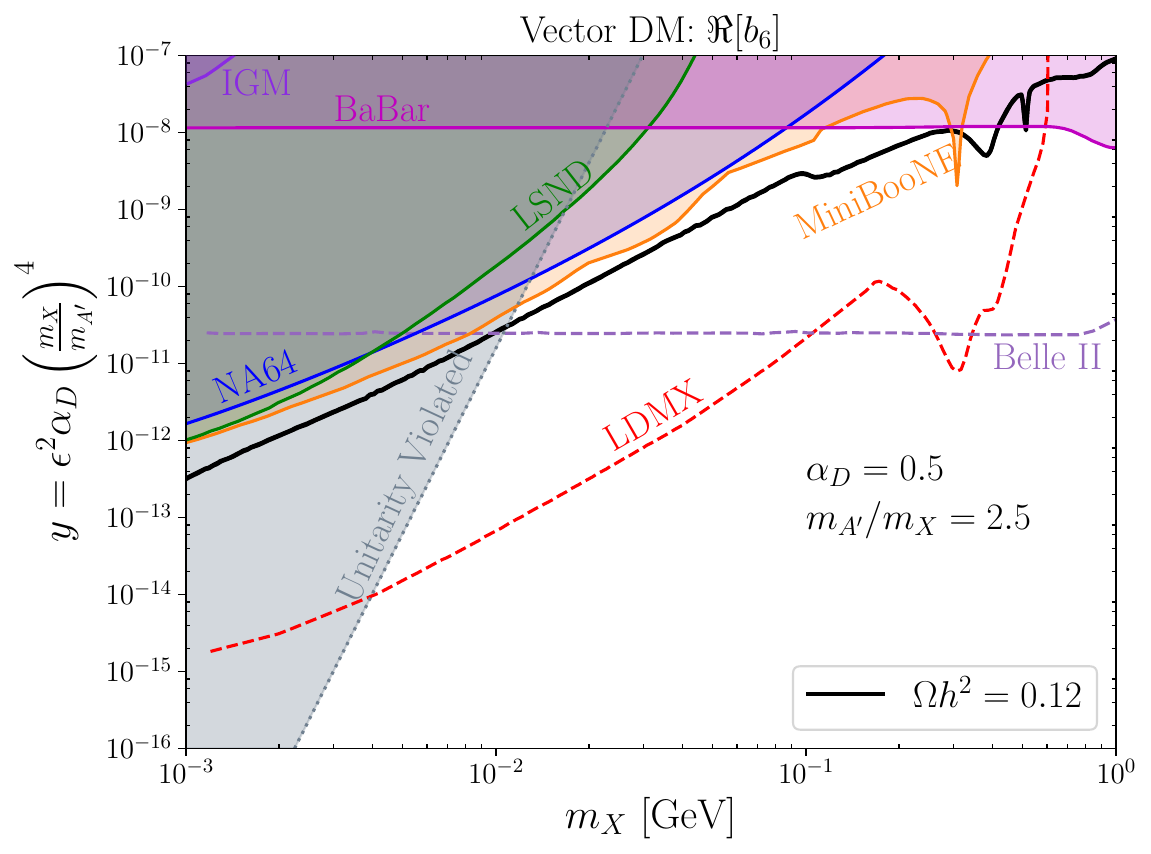}
     \caption{}
     \label{subfig:b6R_ratio2.5}
    \end{subfigure}
    \begin{subfigure}[b]{0.49\textwidth}
    \includegraphics[width=\linewidth]{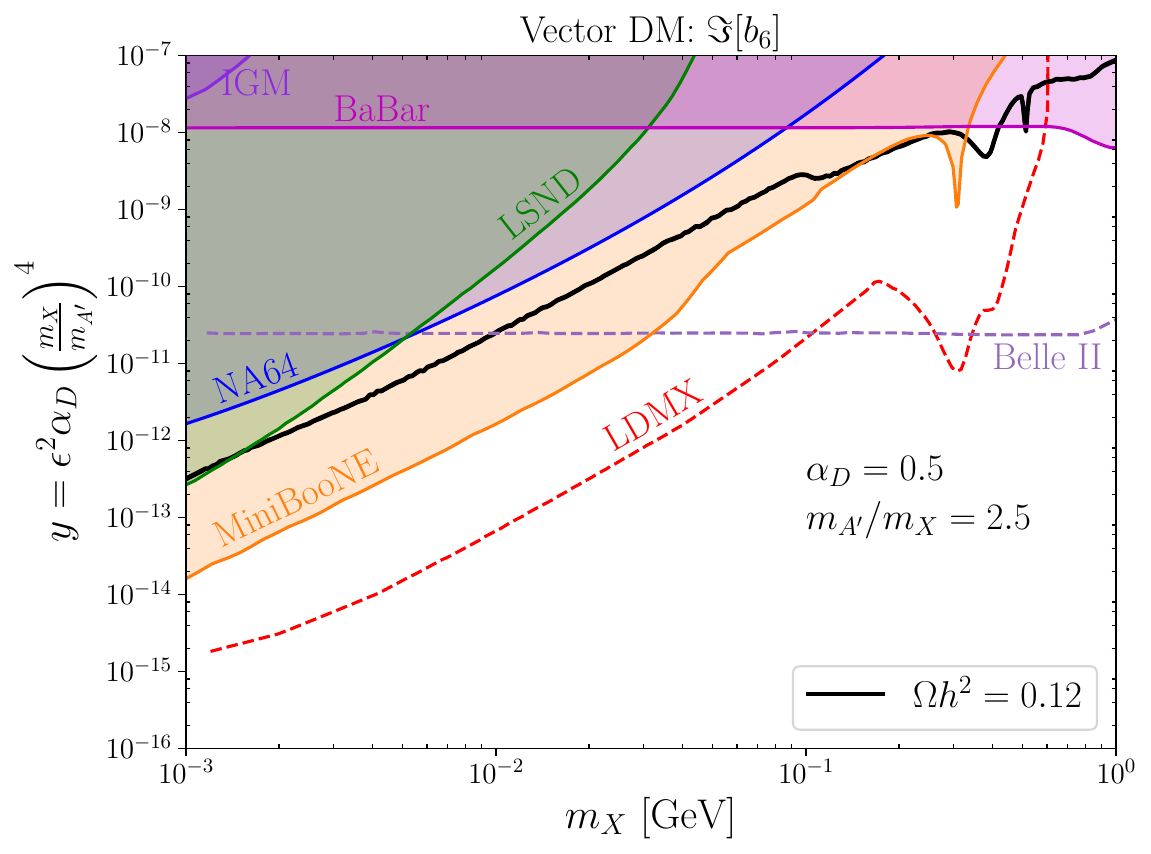}
     \caption{}
     \label{subfig:b6I_ratio2.5}
    \end{subfigure}
    \begin{subfigure}[b]{0.49\textwidth}
    \includegraphics[width=\linewidth]{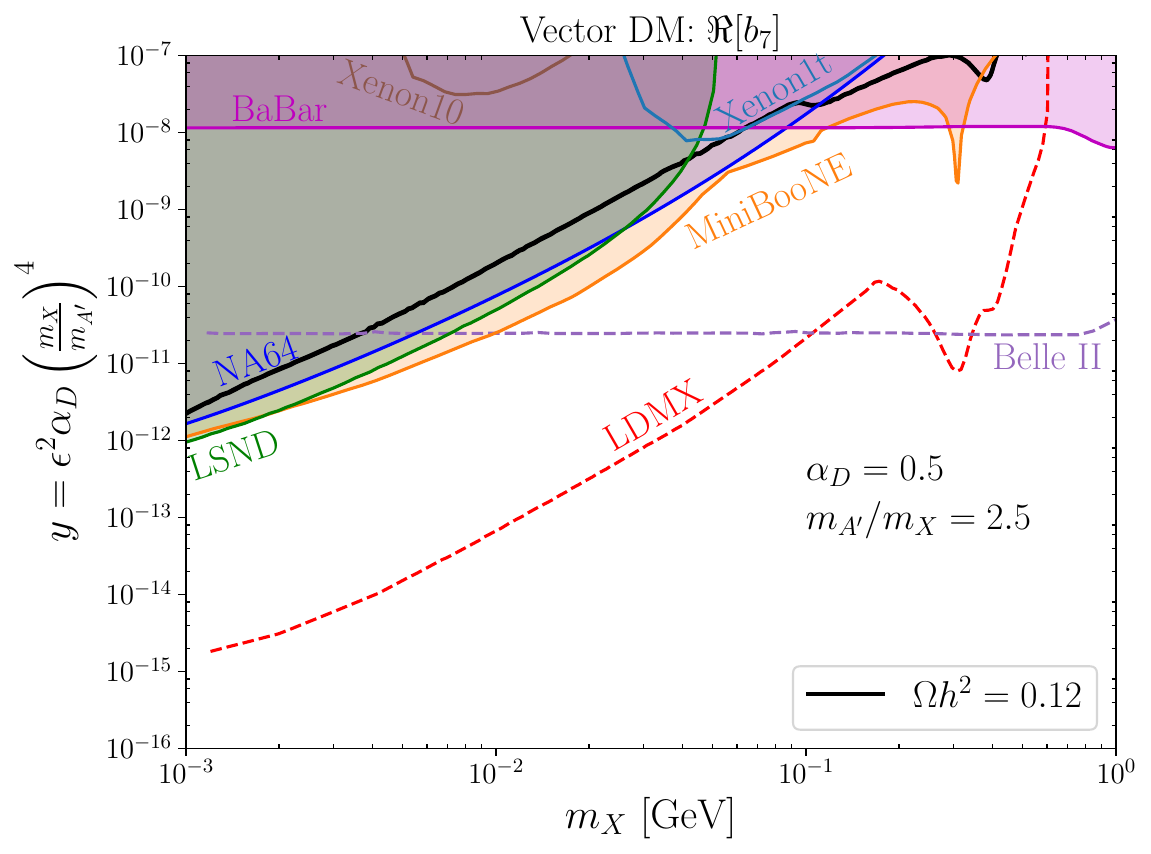}
     \caption{}
     \label{subfig:b7R_ratio2.5}
    \end{subfigure}
    \begin{subfigure}[b]{0.49\textwidth}
    \includegraphics[width=\linewidth]{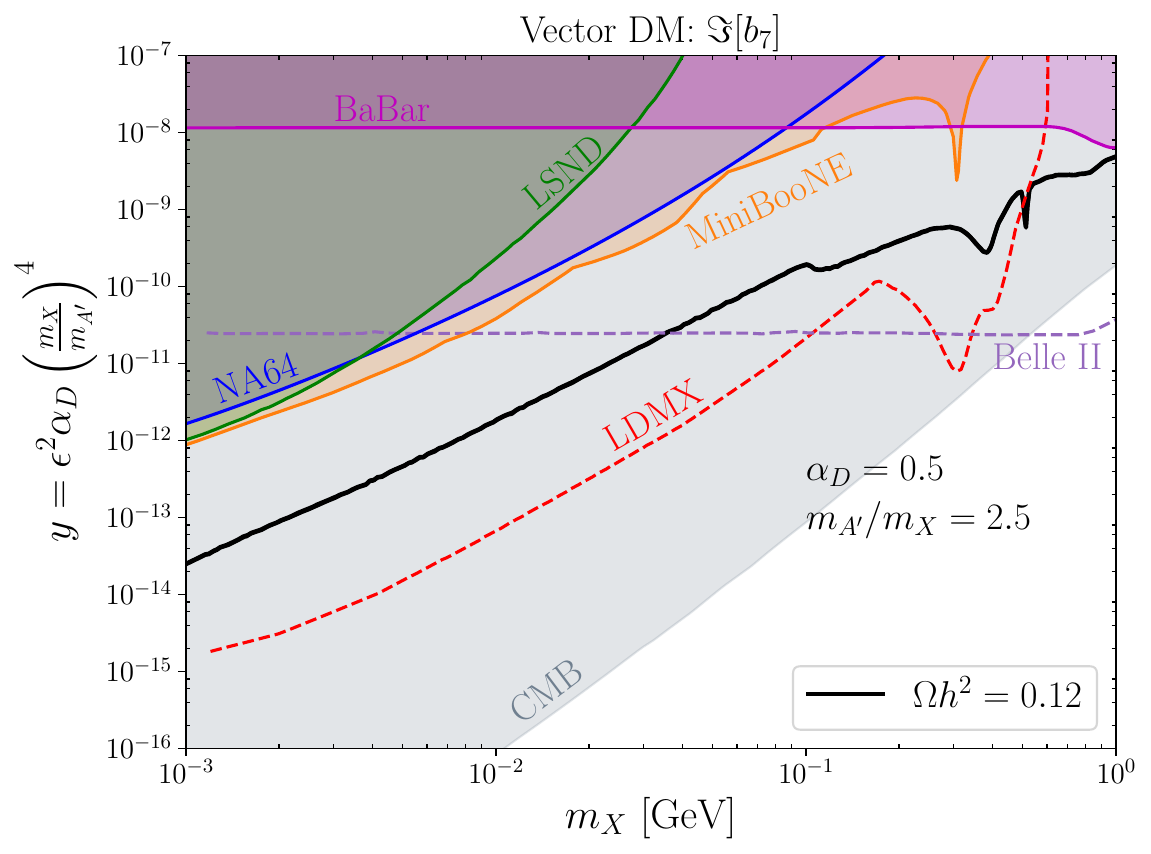}
     \caption{}
     \label{subfig:b7I_ratio2.5}
    \end{subfigure}
    \caption{Same as Fig.~\ref{fig:limits_mratio3}, now for the mass ratio $m_{A'}=2.5~m_{\chi}$.~For this choice of parameters, the $\Re[b_6]$ model predicts the current DM relic abundance, is compatible with observations, and is within reach at LDMX in the mass range between about 10 MeV and 300 MeV.
    }
    \label{fig:limits_mratio2.5}
\end{figure}

\begin{figure}[t]
    \centering
    \includegraphics[width=0.8\textwidth]{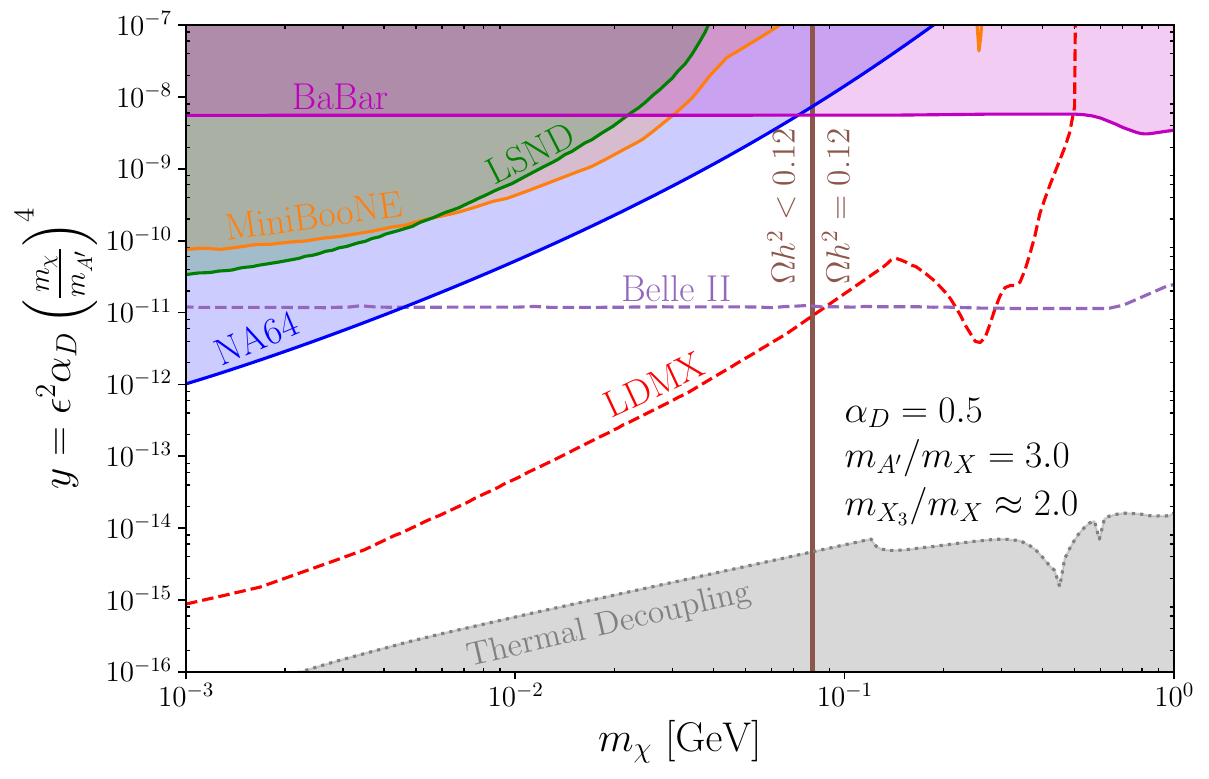}
    \caption{Current and expected exclusion limits on vector SIMP DM in the ($m_X, y$) plane from NA64, MiniBooNE, LSND, and BaBar, as well as from projections for LDMX and Belle II.~The colour code is the one used in previous figures.~The strongest bound on $y$ arises from N64, rather than from MiniBooNE or LSND.~This is due to the fact that the N64  bound on $y$ is model-independent, while the cross section for DM-nucleon and -electron scattering for SIMP DM is suppressed by cancellations between contributions from the $m_{\tilde{Z}'}$ and $m_{\tilde{X}'_3}$ mediators.~In this figure, we assume $\alpha_D=0.5$, $m_{A'}/m_X=3$, and $m_{\tilde{X}_3}/m_X\simeq2$.~Above the grey area, SIMPs are in kinetic equilibrium at chemical decoupling, and can constitute the entire cosmological density of DM on the right of the vertical brown line.}
    \label{fig:SIMPDM}
\end{figure}
\begin{figure}[t]
    \centering
    \includegraphics[width=0.8\textwidth]{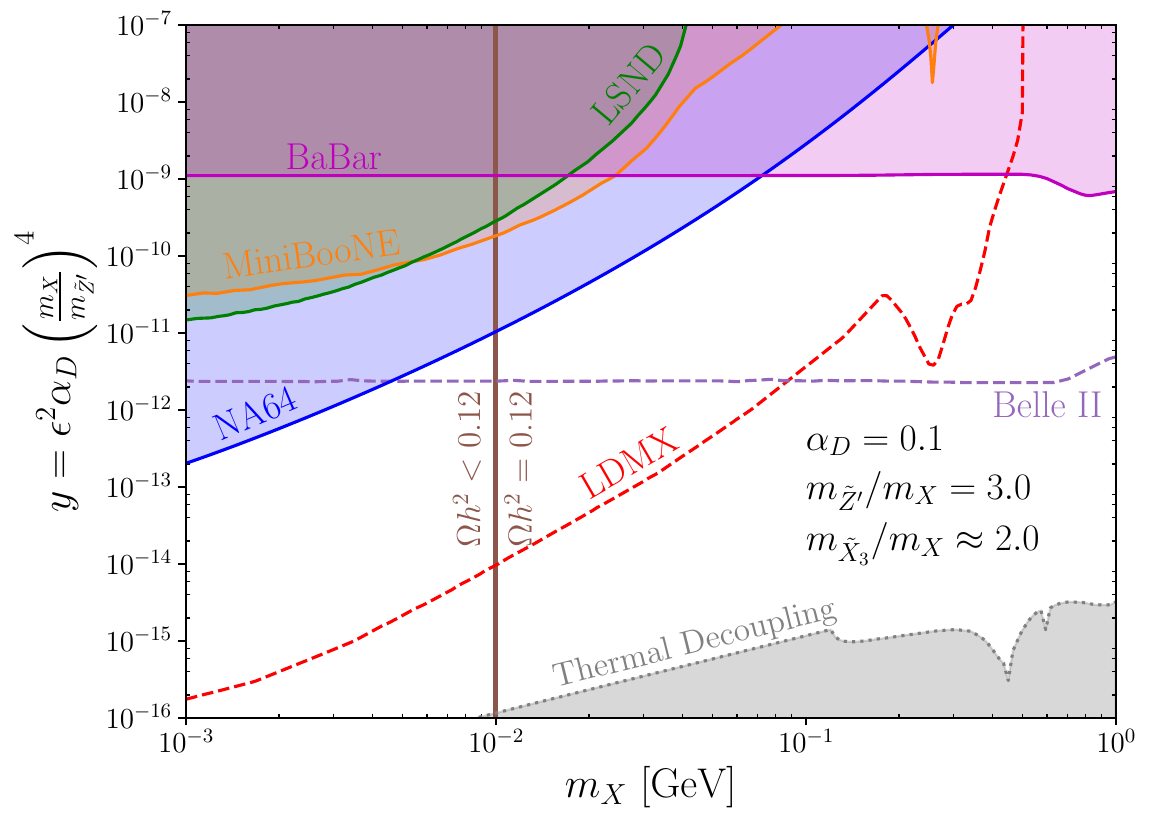}
    \caption{Same as Fig.~\ref{fig:SIMPDM}, now for $\alpha_D=0.1$.}
    \label{fig:SIMPDM2}
\end{figure}

\section{Thermal targets identification}
\label{sec:results}
We now present our spin-1 {\it thermal targets} for DM searches at beam dump and fixed target experiments.~As anticipated, in these regions of parameter space DM is simultaneously thermally produced, not excluded by existing experimental results, and within reach at Belle II or LDMX.~We discuss the simplified models of Sec.~\ref{sec:simp} and the vector SIMP model of Sec.~\ref{sec:non-abelian} separately.

\subsection{Simplified Models} 
Figs. \ref{fig:scalar_b5_limits}, \ref{fig:limits_mratio3}, and \ref{fig:limits_mratio2.5} summarize the  constraints and projections on the $(m_X, y)$ plane that we obtain as explained in Sec.~\ref{sec:bounds} for the simplified models introduced in Sec.~\ref{sec:simp}.~Here,  $y=\epsilon^2 \alpha_D (m_X/m_{A'})^4$ and the black curves on these figures show the contours that are consistent with the observed abundance of DM measured by Planck \cite{Planck:2018vyg}.~The scientific potential of the upcoming experiment LDMX, the area above the red dashed curves of the figures, is significant for sub-GeV DM, since it is projected to probe down to much smaller couplings than previous experiments.

Specifically, the left panels in Fig.~\ref{fig:scalar_b5_limits} show current and expected exclusion limits on the spin-1 model with $b_5\neq 0$ for both $m_{A'}/m_X = 3$ and $m_{A'}/m_X = 2.5$.~The excluded areas in orange, green, blue and purple correspond to the MiniBooNE, LSND, NA64, and BaBar experiments, respectively.~For comparison, we also include the exclusion limits on the $y$ coupling obtained in~\cite{Catena_2023} from the null result reported by the Xenon10 and Xenon1T experiments.~Finally, the expected exclusion limits for LDMX are compared with the expected reach of Belle II.~Independently of the $(m_{A'}/m_X)$ ratio, this model is ruled about by current beam dump experiments.~The right panels in Fig.~\ref{fig:scalar_b5_limits} shows the same set of constraints and projections now for the familiar complex scalar DM model, which we recompute as described in Sec.~\ref{sec:bounds} to calibrate our codes, and for comparison.~Indeed, the complex scalar DM model exhibits the same derivative coupling between DM and the dark photon as in the model with $b_5\neq 0$.~The only difference between the two models is in the Feynman rules for incoming and outgoing DM particles, which in the latter case involves momentum dependent polarisation vectors.~This difference produces a cross section for DM-nucleon scattering that is enhanced by a factor of $(E_{\vec p}/m_X)^2 \gg 1$ relative to the case of the complex scalar DM model, where $E_{\vec p}$ is the relativistic energy of the incoming DM particle in the rest frame of the downstream detector.~As a result, the MiniBooNE constraints on the $b_5$ model are much stronger than in the case of complex scalar DM.

Let us now focus on Fig.~\ref{fig:limits_mratio3}.~This figure shows current and expected exclusion limits on the spin-1 models with $\Re[b_6] \neq 0$ and $\Im[b_6]\neq 0$ (top left and top right panels) as well as on the models with $\Re[b_7] \neq 0$ and $\Im[b_7] \neq 0$ (bottom left and bottom right panels).~In all panels we assume $m_{A'}/m_X = 3$, while the colour code is the same as in Fig.~\ref{fig:scalar_b5_limits}.~Current beam dump experiments, LSND and MiniBooNE  (green and orange shaded regions respectively), and BaBar (pink shaded regions) are able to rule out large parts of the $(m_X,y)$ plane for the $\Re[b_7]$, $\Im[b_6]$ and $\Im[b_7]$ models, while leaving the $\Re[b_6] \neq 0$ spin-1 model still compatible with observations in the mass range between approximately 40 MeV and 200 MeV.  
Remarkably, the constraints from MiniBooNE on $y$ are much stronger in the case of the $\Im[b_6]$ model than for the $\Re[b_6]$ model.~This is due to the fact that for the $\Im[b_6]$ model the cross section for DM-nucleon scattering is enhanced by a $(E_{\vec p}/m_X)^2 \gg 1$ factor relative to the analogous cross section for the $\Re[b_6]$ model.~Here $E_{\vec p}$ is the energy of the DM particle in the downstream detector rest frame.~As anticipated above, a similar enhancement is also present in the case of the $b_5\neq 0$ spin-1 model.~Finally, in the case of the $\Re[b_6]$ model, top left panel in Fig.~\ref{fig:limits_mratio3}, we also report the unitary bound on $y$ arising from the helicity amplitude for DM-electron scattering, which is one of the processes directly entering the calculation of the constraints in Fig.~\ref{fig:limits_mratio3}.

It should also be noticed that for the $\Im[b_7]$ spin-1 model, Figs. \ref{subfig:b7I_ratio3} and \ref{subfig:b7I_ratio2.5}, the DM annihilation cross section is s-wave dominant in contrast to being p-wave dominant as in the other models we consider, leading to a strong constraint from CMB measurements on this scenario.

Fig.~\ref{fig:limits_mratio2.5} shows current and expected exclusion limits on the same models as in Fig.~\ref{fig:limits_mratio3}, now assuming the different mass ratio, $m_{A'}/m_X = 2.5$.~While this leaves are conclusions qualitatively unchanged, the range of DM masses that are compatible with all observations in the case of the $\Re[b_6]$ model is now significantly broader, varying from about 10 MeV to 300 MeV. 

\subsection{Non-abelian SIMPs}
 We now turn our attention to the case of vector SIMP DM.~We start by reviewing some of the features of the model, and its differences with the simplified models.~In the SIMP model, the DM relic abundance is set by the freeze-out of $3\rightarrow 2$ processes and forbidden annihilations, and is thus independent of the coupling $y$.~At the same time, for the thermal production of SIMP DM to work, DM has to be in kinetic equilibrium in the early universe.~As anticipated, this occurs via effective DM-$\tilde{X}_3$ mediator scattering processes combined with in-equilibrium $\tilde{X}_3$ decays into SM particles.~Since the decay rate of $\tilde{X}_3$ depends on $\epsilon$~\cite{Choi:2019zeb}, only above a certain critical value for $y$ vector SIMP DM can effectively be thermally produced in the right amount.~As described in Sec.~\ref{sec:theory}, we calculate this lower bound by using Eq.~(\ref{eq:kin}).~The result of this calculation defines a region in the parameter space of the model above which SIMPs are in kinetic equilibrium at their chemical decoupling, and the freeze-out mechanics can in principle successfully predict the present DM cosmological density.~Whether or not the whole cosmological DM abundance is in the form of vector SIMPs depends on the choice of $\alpha_D$.~For example, for $\alpha_D=0.1$ vector SIMPs can constitute the entire DM in our universe for masses above about 10 MeV, whereas for $\alpha_D=0.5$ this is only possible for masses above about 80 MeV.~Here and in the figures below, we assume the benchmark value for the mixing angle $\sin(2 \theta'_X)=-0.1$.

Keeping these general considerations on SIMP DM in mind, we are now ready to compare the predictions of the vector SIMP DM model with the experimental and theoretical constraints of Sec.~\ref{sec:bounds}.~Fig.~\ref{fig:SIMPDM} shows the current and expected exclusion limits on the $y$ coupling as a function of the DM mass that we obtain from a reanalysis of data collected at NA64, MiniBooNE, LSND, and BaBar, as well as from projections for LDMX and Belle II.~We obtain these exclusion limits by comparing theory and observations as described in Sec.~\ref{sec:bounds}, while the colour code in the figure is the one of the figures in the previous sections.~Remarkably, in the case of SIMP DM the strongest bound on $y$ arises from NA64, rather than from MiniBooNE or LSND.~This is due to the fact that the NA64 bound on $y$ is model-independent, while the cross section for DM-nucleon and -electron scattering for SIMP DM is suppressed by cancellations between contributions from the $m_{\tilde{Z}'}$ and $m_{\tilde{X}'_3}$ mediators.~In Fig.~\ref{fig:SIMPDM}, we assume $\alpha_D=0.5$, $m_{A'}/m_X=3$, and $m_{\tilde{X}_3}/m_X\simeq2$.~Fig.~\ref{fig:SIMPDM2} reports the results of an analogous analysis where we assume a different combination of parameters, namely $\alpha_D=0.1$, $m_{A'}/m_X=3$, and $m_{\tilde{X}_3}/m_X\simeq 2$.~The main difference between the analyses is the range of masses for which the whole cosmological DM abundance is in the form of vector SIMPs.~As explained above, this difference is due to the different assumptions made for $\alpha_D$ in the two figures.

\section{Conclusion}
\label{sec:conclusion}

In this analysis, we extended the current landscape of sub-GeV DM models considered in the context of various experiments such as MiniBooNE, LSND and LDMX to a set of models for spin-1 DM, including a general family of simplified models (involving one DM particle and one mediator -- the dark photon) and an ultraviolet complete model based on a non-abelian gauge group (now including two mediators and an extended Higgs sector) where DM is a vector SIMP.~For each of these models, we calculated the DM relic density, the expected number of signal events in beam dump experiments such as LSND and MiniBooNE, the rate of energy injection in the early universe thermal bath and in the IGM, as well as the helicity amplitudes for forward processes subject to the unitary bound.~We then compared these predictions with a number of different experimental results from Planck, CMB observations, direct detection experiments (Xenon10 and Xenon1T), data on the IGM temperature from Lyman alpha observations, LSND, MiniBooNE, NA64, and BaBar and with available projections from LDMX and Belle II.~Through this comparison, we identified the regions in the parameter space of the models considered in this work where DM is simultaneously thermally produced, compatible with present observations, and within reach at Belle II and, in particular, at LDMX.

We found that the simplified models for spin-1 DM investigated in our analysis are strongly constrained by LSND and MiniBooNE, as well as from bounds on the unitarity of the S-matrix.~The only model not already excluded by these experimental and theoretical constraints is the one characterised by the coupling constants $h_3$ and $\Re[b_6]$.~For a dark photon to DM mass ratio of 3 (2.5), this model is compatible with current observations, within reach at LDMX and admits a thermal DM candidate for DM masses in a window between about 40 (10) MeV and 200 (300) MeV.~In this mass range, the model has a relic density contour lying very close to current 90\% C.L. exclusion limits from beam dump experiments in the ($m_X, y$) plane, and will thus be conclusively probed (i.e. excluded or discovered) in the first LDMX run.

At the same time, we found that the vector SIMP model explored in this work admits thermal DM candidates that are not ruled out by beam dump experiments and within reach at LDMX in a wide region of the underlying parameter space.~The model features a DM production mechanism that is complementary to the freeze-out of DM pair annihilations into SM particles of simplified models, and is based on the interplay of $3 \rightarrow 2$ processes and forbidden annihilations.~It also exhibits a lower bound on the DM particle mass arising from the relic density constraint.~The larger $\alpha_D$, the larger the minimum admissible DM particle mass.

Ultimately, our investigation bridges a gap in the current knowledge of sub-GeV DM by providing the DM community with new sub-GeV spin-1 thermal targets lying in the experimentally accessible region of next-generation beam dump and fixed target experiments such as LDMX.

\acknowledgments 
We would like to thank Felix Kahlhoefer and Chris Chang for pointing out the importance of the unitarity bound on simplified models for spin-1 DM.~We would also like to thank Avik Banerjee and Gabriele Ferretti for drawing our attention to the work by Choi {\it et al.}~\cite{Choi:2019zeb} on vector SIMP DM.~In addition, we greatly appreciate the important discussions we had with Patrick deNiverville on beam dump experiment simulations, in particular his software BdNMC.~The research contained in this article was performed within the Knut and Alice Wallenberg project grant Light Dark Matter (Dnr.~KAW 2019.0080).~We would like to thank all participants in the project for valuable discussions on sub-GeV DM and the physics reach of LDMX during our weekly collaboration meetings.~R.C. acknowledges support from individual research grants from the Swedish Research Council, Dnr. 2018-05029 and Dnr. 2022-04299.

\appendix

\section{Cross sections for relativistic dark matter-nucleon scattering}
\label{sec:sigmasimpN}
In this appendix, we provide analytic expressions for the differential cross sections for relativistic DM-nucleon scattering for the spin-1 DM models of Sec.~\ref{sec:theory}.~As anticipated, we obtain these cross sections by implementing the models of Sec.~\ref{sec:theory} in {\sffamily FeynRules}~\cite{Alloul:2013bka} and then using {\sffamily CalcHEP}~\cite{Belyaev:2012qa} to generate analytic expressions for the squared modulus of the spin-averaged scattering amplitude.~We finally validate the outcome of this symbolic calculation through direct analytical calculations of a subset of selected cross sections.~In the laboratory frame, we find
\begin{align}
\frac{{\rm d} \sigma_{NX}(E_{\vec p}, E_{\vec p'})}{{\rm d} E_{\vec p'}} = \frac{1}{32 \pi m_N (E_{\vec p}^2-m_X^2)} \overline{|\mathcal{M}_{NX}(E_{\vec p}, E_{\vec p'})|^2} \,,
\label{eq:sigma_N}
\end{align}
where $E_{\vec p}$ ($E_{\vec p'}$) is the initial (final) DM particle energy and the squared scattering amplitude is given by
\begin{align}
 \overline{|\mathcal{M}_{NX}(E_{\vec p}, E_{\vec p'})|^2} &= \frac{\eta^2\left[ F_1^2 \mathcal{A}_\eta(E_{\vec p}, E_{\vec p'}) +  F_2^2 \mathcal{B}_\eta(E_{\vec p}, E_{\vec p'}) + F_1F_2\mathcal{C}_\eta(E_{\vec p}, E_{\vec p'})\right]}{3 m_X^4 \left[2 m_N (E_{\vec p}- E_{\vec p'})+m_{A'}^2\right]^2}  \,.
 \label{eq:M2}
\end{align}
Here, a dependence of the nucleon form factors $F_1(q^2)$ and $F_2(q^2)$ on the momentum transfer $q=p-p'$ is understood.~The three functions, $\mathcal{A}_{\eta}(E_{\vec p}, E_{\vec p'})$, $\mathcal{B}_{\eta}(E_{\vec p}, E_{\vec p'})$ and $\mathcal{C}_{\eta}(E_{\vec p}, E_{\vec p'})$ in Eq.~(\ref{eq:M2}) are model dependent and have dimension of mass to the eighth power, so that  $\overline{|\mathcal{M}_{NX}(E_{\vec p}, E_{\vec p'})|^2}$  is dimensionless.~$\eta$ is the coupling constant characterising the underlying DM model.~Below, we specify $\mathcal{A}_{\eta}(E_{\vec p}, E_{\vec p'})$, $\mathcal{B}_{\eta}(E_{\vec p}, E_{\vec p'})$ and $\mathcal{C}_{\eta}(E_{\vec p}, E_{\vec p'})$ for different choices of $\eta$.

\begin{itemize}
\item For $\eta=b_5$, we find
\begin{align}
\mathcal{A}_{b_5}(E_{\vec p}, E_{\vec p'})&= 8  m_N \left[E_{\vec p'} \left(2 E_{\vec p} m_N+m_X^2\right)-E_{\vec p} m_X^2\right] 
\sum_s\epsilon^s_{\mu} \epsilon^{s \mu *} \,,\nonumber\\
\mathcal{B}_{b_5}(E_{\vec p}, E_{\vec p'}) &= 2  m_N \left(E_{\vec p}-E_{\vec p'}\right) \left[E_{\vec p}^2+2 (E_{\vec p}+m_N)
   E_{\vec p'} +E_{\vec p'}^2-2 E_{\vec p} m_N-4 m_X^2\right] \nonumber\\
   &\times \sum_s\epsilon^s_{\mu} \epsilon^{s \mu *} \,,\nonumber\\
\mathcal{C}_{b_5}(E_{\vec p}, E_{\vec p'})&=-8 m_N \left(E_{\vec p}-E_{\vec p'}\right) \left(-m_N E_{\vec p'}+E_{\vec p} m_N+2
   m_X^2\right)\sum_s\epsilon^s_{\mu} \epsilon^{s \mu *} \,,
\end{align}
where the sum over spin configurations of the product of DM polarisation vectors is given by
\begin{align}
\sum_s\epsilon^s_{\mu} \epsilon^{s \mu *}  = \left\{E_{\vec p}^2 m_N^2+m_N E_{\vec p'} \left[m_N E_{\vec p'}-2 \left(E_{\vec p} m_N+m_X^2\right)\right]+2
   E_{\vec p} m_N m_X^2+3 m_X^4\right\} \,.
\end{align}
\item For $\eta=\Re(b_6)$, we obtain
\begin{align}
\mathcal{A}_{\Re(b_6)}(E_{\vec p}, E_{\vec p'})&= 8  m_N^2 m_X^2\left(E_{\vec p}-E_{\vec p'}\right) \left[E_{\vec p'} \left(m_N
   E_{\vec p'}+m_N^2-m_X^2\right)+m_X^2 (E_{\vec p}-2 m_N)\right. \nonumber \\
   &\left. +E_{\vec p} m_N (E_{\vec p}-m_N)\right] \,,\nonumber\\
\mathcal{B}_{\Re(b_6)}(E_{\vec p}, E_{\vec p'})&=4 m_N m_X^2\left(E_{\vec p}-E_{\vec p'}\right)^2 \left[E_{\vec p'} \left(2 E_{\vec p}
   m_N-m_N^2+m_X^2\right)\right.\nonumber\\
   &\left. +E_{\vec p} (m_N-m_X) (m_N+m_X)+2 m_N m_X^2\right] \,, \nonumber\\
\mathcal{C}_{\Re(b_6)}(E_{\vec p}, E_{\vec p'})&=16 m_N^2m_X^2 \left(E_{\vec p}-E_{\vec p'}\right)^2 \left(-m_N E_{\vec p'}+E_{\vec p} m_N+2 m_X^2\right) \,.
\end{align}
\item For $\eta=\Im(b_6)$, the model dependent functions in Eq.~(\ref{eq:M2}) can explicitly be written as follows
\begin{align}
\mathcal{A}_{\Im(b_6)}(E_{\vec p}, E_{\vec p'})&=8 m_N^2 \left(E_{\vec p}-E_{\vec p'}\right) \left\{E_{\vec p'} \left[2 E_{\vec p}^2 m_N^2-2 E_{\vec p}
   m_N^2 E_{\vec p'}+m_N m_X^2 (2 E_{\vec p}+m_N)\right.\right. \nonumber\\
   &\left.\left. -m_X^4\right]-E_{\vec p} m_N^2 m_X^2+m_X^4
   (E_{\vec p}-2 m_N)\right\} \,, \nonumber\\
\mathcal{B}_{\Im(b_6)}(E_{\vec p}, E_{\vec p'})&=2 m_N \left(E_{\vec p}-E_{\vec p'}\right)^2 \left\{E_{\vec p}^2 m_N^2 (E_{\vec p}-2 m_N) \right. \nonumber\\  &\left. +E_{\vec p'}
   \left[-m_N^2 E_{\vec p'} \left(E_{\vec p'}+E_{\vec p}+2 m_N\right)+E_{\vec p} m_N^2 (E_{\vec p}+4 m_N) \right.\right. \nonumber\\
   &\left.\left.+2 m_N
   m_X^2 (2 E_{\vec p}+m_N)+2 m_X^4\right]-2 E_{\vec p} m_N^2 m_X^2-2 m_X^4 (E_{\vec p}-2 m_N)\right\}  \,, \nonumber\\
\mathcal{C}_{\Im(b_6)}(E_{\vec p}, E_{\vec p'})&=-8  m_N^2 \left(E_{\vec p}-E_{\vec p'}\right)^2 \left[E_{\vec p}^2 m_N^2+m_N^2 E_{\vec p'}
   \left(E_{\vec p'}-2 E_{\vec p}\right)-4 m_X^4\right] \,.
\end{align}

\item For $\eta=\Re(b_7)$, we find
\begin{align}
\mathcal{A}_{\Re(b_7)}(E_{\vec p}, E_{\vec p'})&=8 m_N m_X^2\left(-m_N E_{\vec p'}+E_{\vec p} m_N+2 m_X^2\right) \left[E_{\vec p'} \left(m_N
   E_{\vec p'}+m_N^2-m_X^2\right) \right. \nonumber\\
   & \left.+m_X^2 (E_{\vec p}-2 m_N)+E_{\vec p} m_N (E_{\vec p}-m_N)\right] \,, \nonumber\\
\mathcal{B}_{\Re(b_7)}(E_{\vec p}, E_{\vec p'})&=4 m_X^2 \left(E_{\vec p}-E_{\vec p'}\right) \left(-m_N E_{\vec p'}+E_{\vec p} m_N+2 m_X^2\right) \nonumber\\
&\times   \left[E_{\vec p'} \left(2 E_{\vec p} m_N-m_N^2+m_X^2\right) +E_{\vec p} (m_N-m_X) (m_N+m_X)\right. \nonumber\\
&\left.+2 m_N m_X^2\right] \,, \nonumber\\
\mathcal{C}_{\Re(b_7)}(E_{\vec p}, E_{\vec p'})&= 16 m_N m_X^2\left(E_{\vec p}-E_{\vec p'}\right) \left(-m_N E_{\vec p'}+E_{\vec p} m_N+2
   m_X^2\right)^2 \,.
\end{align}
\item Finally, for $\eta=\Im(b_7)$, we obtain
\begin{align}
\mathcal{A}_{\Im(b_7)}(E_{\vec p}, E_{\vec p'})&= 8  m_N^2 m_X^2\left(E_{\vec p}-E_{\vec p'}\right) \left[E_{\vec p}^2 m_N+E_{\vec p'} \left(m_N
   E_{\vec p'}+m_N^2+m_X^2\right) \right. \nonumber\\
   &\left.-E_{\vec p} \left(m_N^2+m_X^2\right)\right] \,,\nonumber\\
\mathcal{B}_{\Im(b_7)}(E_{\vec p}, E_{\vec p'})&= 4 m_N^2m_X^2 \left(E_{\vec p}-E_{\vec p'}\right)^2 \left[(2 E_{\vec p}-m_N) E_{\vec p'}+E_{\vec p} m_N-2
   m_X^2\right]\,, \nonumber\\
\mathcal{C}_{\Im(b_7)}(E_{\vec p}, E_{\vec p'})&= 16 m_N^2 m_X^2\left(E_{\vec p}-E_{\vec p'}\right)^2 \left(-m_N E_{\vec p'}+E_{\vec p}
   m_N-m_X^2\right) \,.
\end{align}
\end{itemize}

\section{Cross sections for relativistic dark matter-electron scattering}
\label{sec:sigmasimp}
In the laboratory frame, the differential cross section for DM-electron scattering can be written as follows
\begin{align}
\frac{{\rm d} \sigma_{eX}(E_{\vec p}, E_{\vec k'})}{{\rm d} E_{\vec k'}} = \frac{1}{32 \pi m_e (E_{\vec p}^2-m_X^2)} \overline{|\mathcal{M}_{eX}(E_{\vec p}, E_{\vec k'})|^2} \,,
\label{eq:sigma_e}
\end{align}
where $E_{\vec k'}$ is the final state electron energy.~As for the case of DM-nucleon scattering, we evaluate Eq.~(\ref{eq:sigma_e}) by the combined use of {\sffamily FeynRules}~\cite{Alloul:2013bka}, {\sffamily CalcHEP}~\cite{Belyaev:2012qa} and analytical calculations for validation.~Here, we express the squared modulus of the scattering amplitude in Eq.~(\ref{eq:sigma_e}) as
\begin{align}
\overline{|\mathcal{M}_{eX}(E_{\vec p}, E_{\vec k'})|^2} = \frac{\eta^2 h_3^2\,\mathcal{D}_\eta(E_{\vec p},E_{\vec k'})}{3 m_X^4 \left[ 2m_e(E_{\vec k'}-m_e)+m_{A'}^2\right]^2} \,.
\end{align}
Below, we specify the model dependent function $\mathcal{D}_\eta(E_{\vec p},E_{\vec k'})$ for different choices of coupling constant $\eta$:
\begin{align}
\mathcal{D}_{b_5}(E_{\vec p},E_{\vec k'}) &=
8m_e \left[m_e^2 (E_{\vec k'}-m_e)^2+2 m_e m_X^2 (E_{\vec k'}-m_e)+3 m_X^4\right]\nonumber\\
&\times \left[2
   E_{\vec p} m_e (-E_{\vec k'}+E_{\vec p}+m_e)+m_X^2 (m_e-E_{\vec k'})\right]
 \,,
\nonumber\\
\mathcal{D}_{\Re(b_6)}(E_{\vec p},E_{\vec k'}) &= 8 m_e^2 m_X^2 (E_{\vec k'}-m_e) \left\{m_e \left[E_{\vec k'}^2-E_{\vec k'} (2
   E_{\vec p}+3m_e)+2 \left(E_{\vec p}^2+E_{\vec p}m_e+m_e^2\right)\right]\right.\nonumber\\
   &\left.+m_X^2 (E_{\vec k'}-3 m_e)\right\} \,,
\nonumber\\
\mathcal{D}_{\Im(b_6)}(E_{\vec p},E_{\vec k'}) &= -8 m_e^2 (E_{\vec k'}-m_e) \left\{-m_e m_X^2 \left[-E_{\vec k'} (2
   E_{\vec p}+m_e)+2 E_{\vec p}^2+2 E_{\vec p}m_e+m_e^2\right]\right. \nonumber\\
   &\left.+2 E_{\vec p}m_e^2 (m_e-E_{\vec k'})
   (-E_{\vec k'}+E_{\vec p}+m_e)-m_X^4 (E_{\vec k'}-3m_e)\right\} \,,
\nonumber\\
\mathcal{D}_{\Re(b_7)}(E_{\vec p},E_{\vec k'}) &= -8 m_e m_X^2 \left(-E_{\vec k'}m_e+m_e^2-2 m_X^2\right) \left\{m_e
   \left[E_{\vec k'}^2-E_{\vec k'} (2 E_{\vec p}+3m_e)\right.\right. \nonumber\\  
   &\left.\left.+2 \left(E_{\vec p}^2+E_{\vec p}
  m_e+m_e^2\right)\right]+m_X^2 (E_{\vec k'}-3m_e)\right\} \,,
\nonumber\\
\mathcal{D}_{\Im(b_7)}(E_{\vec p},E_{\vec k'}) &= 8 m_e^2 m_X^2(E_{\vec k'}-m_e) \left\{m_e \left[E_{\vec k'}^2-E_{\vec k'} (2 E_{\vec p}+3 m_e)+2
   \left(E_{\vec p}^2+E_{\vec p} m_e+m_e^2\right)\right] \right.\nonumber\\
   &\left.+m_X^2 (m_e-E_{\vec k'})\right\} \,.
\end{align}

\section{Scattering cross sections for SIMP DM}
\label{app:non-abelian}
For the DM-electron scattering cross section for the non-abelian SIMP model, we find the expression
\begin{align}
\frac{{\rm d}\sigma}{{\rm d} E_e} &=
\frac{1}{192 \pi m_X^4}  
\frac{\sin^2(2\theta'_X) e^2 \epsilon ^2  g_X^2}{E_{\vec p}^2-m_X^2} \Big\{ \mathscr{A}(E_e) \left[ E_{\vec p}^2 - (E_e - m_e) E_{\vec p} \right] - \mathscr{B}(E_e)\Big\} \mathscr{C}(E_e) \,,
\label{eq:sigmaeSIMP}
\end{align}
where we collected terms depending on different powers of $E_{\vec p}$, and introduced the two coefficients 
\begin{align}
\mathscr{A}(E_e)&= 2(E_e - m_e)^2 m_e^3 + 10 (E_e - m_e) m_e^2 m_X^2 + 24 m_e m_X^4\,,
\end{align}
and
\begin{align}
\mathscr{B}(E_e)&= (E_e-m_e) m_X^2 \left[ (E_e-m_e) m_e^3 + (3 E_e -m_e) m_e m_X^2 + 12 m_X^4 \right] \,.
\end{align}
The overall factor $\mathscr{C}(E_e)$ arises from the propagators due to $\tilde{Z}'$ and $\tilde{X}_3$ exchange, and it is given by
\begin{align}
\mathscr{C}(E_e)= \left(\frac{1}{2 E_e m_e-2 m_e^2+m_{\tilde{Z}'}^2}-\frac{1}{2 E_e m_e-2 m_e^2+m_{\tilde{X}_3}^2}\right)^2\,.
\end{align}
In the non-relativistic limit, Eq.~(\ref{eq:sigmaeSIMP}) reduces to Eq.~(66) from~\cite{Choi:2019zeb} if integrated from 0 to $2 \mu^2 v^2/m_e$, where $\mu$ is the DM-electron reduced mass, while $v$ is the DM-electron relative velocity.

For the DM-nucleon scattering cross section for the non-abelian SIMP model, we find the expression
\begin{align}
\frac{{\rm d} \sigma}{{\rm d} E_{\vec p'}} &= \frac{1}{192 \pi m_X^4}\frac{\sin^2(2\theta'_X) e^2 \epsilon ^2  g_X^2}{E_{\vec p}^2-m_X^2} 
\left[ F_1^2 \mathcal{A}_1(E_{\vec p}, E_{\vec p'}) +  F_2^2 \mathcal{A}_2(E_{\vec p}, E_{\vec p'}) + 2 F_1F_2\mathcal{A}_{12}(E_{\vec p}, E_{\vec p'})\right] \nonumber\\
&\times \frac{\mathscr{D}(E_{\vec p},E_{\vec p'})}{4}\,,
\label{eq:sigma_N}
\end{align}
where
\begin{align}
\mathcal{A}_1(E_{\vec p}, E_{\vec p'})&=4 \Big\{2 E_{\vec p} m_N^3 E_{\vec p'}^3+2
   E_{\vec p'} \Big[E_{\vec p}^3 m_N^3+E_{\vec p} m_N^2 m_X^2 (5
   E_{\vec p}+m_N) \nonumber\\
   &+m_N m_X^4 (15 E_{\vec p}+m_N)+6 m_X^6 \Big]
   \nonumber\\
   &-m_N
   E_{\vec p'}^2 \Big[4 E_{\vec p}^2 m_N^2+m_N m_X^2 (10
   E_{\vec p}+m_N)+3 m_X^4\Big]
   \nonumber\\
   &-E_{\vec p} m_X^2 \Big[E_{\vec p} m_N^3+m_N
   m_X^2 (3 E_{\vec p}+2 m_N)+12 m_X^4\Big]\Big\} \,,
\end{align}

\begin{align}
\mathcal{A}_2(E_{\vec p}, E_{\vec p'})&=\left(E_{\vec p}-E_{\vec p'}\right) \Big\{E_{\vec p}^3 m_N^2
   (E_{\vec p}-2 m_N)+2 E_{\vec p}^2 m_N m_X^2 (2 E_{\vec p}-5 m_N)\nonumber\\
   &+E_{\vec p'} \Big[6
   E_{\vec p}^2 m_N^3
   +E_{\vec p'} \Big(m_N E_{\vec p'} (m_N E_{\vec p'}+2 m_N^2-4
   m_X^2)-2 E_{\vec p} m_N^2 (E_{\vec p}+3 m_N)\nonumber\\
   &-2 m_N m_X^2 (4
   E_{\vec p}+5 m_N)+10 m_X^4\Big)+4 m_X^4 (7 E_{\vec p}+9 m_N)\nonumber\\
   &+4 E_{\vec p}
   m_N m_X^2 (2 E_{\vec p}+5 m_N)\Big]+2 E_{\vec p} m_X^4 (5 E_{\vec p}-18
   m_N)-48 m_X^6\Big\} \,,
\end{align}

\begin{align}
\mathcal{A}_{12}(E_{\vec p}, E_{\vec p'})&=-2\left(E_{\vec p}-E_{\vec p'}\right) \left(-m_N E_{\vec p'}+E_{\vec p} m_N+2 m_X^2\right) \Big\{E_{\vec p}^2 m_N^2+m_N E_{\vec p'}
   \big[m_N E_{\vec p'} \nonumber\\
   &-2 \left(E_{\vec p} m_N+m_X^2\right)\Big]+2 E_{\vec p} m_N
   m_X^2+12 m_X^4\Big\} \,,
\end{align}
and
\begin{align}
\mathscr{D}(E_{\vec p},E_{\vec p'})= \left(\frac{1}{m_{\tilde{Z}'}^2+2 m_N(E_{\vec p} -E_{\vec p'})}-\frac{1}{m_{\tilde{X}_3}+2 m_N(E_{\vec p} -E_{\vec p'})}\right)^2\,.  
\end{align}

\bibliographystyle{JHEP}
\bibliography{vecdm}

\providecommand{\href}[2]{#2}\begingroup\raggedright\begin{thebibliography}{10}

\bibitem{Cooley:2022ufh}
J.~Cooley et~al., \emph{{Report of the Topical Group on Particle Dark Matter
  for Snowmass 2021}},  \href{https://arxiv.org/abs/2209.07426}{{\ttfamily
  2209.07426}}.

\bibitem{Battaglieri:2017aum}
M.~Battaglieri et~al., \emph{{US Cosmic Visions: New Ideas in Dark Matter 2017;
  Community Report}}, {\emph{FERMILAB-CONF-17-282-AE-PPD-T} (2017) }
  [\href{https://arxiv.org/abs/1707.04591}{{\ttfamily 1707.04591}}].

\bibitem{Mitridate:2022tnv}
A.~Mitridate, T.~Trickle, Z.~Zhang and K.~M. Zurek, \emph{{Snowmass white
  paper: Light dark matter direct detection at the interface with condensed
  matter physics}},
  \href{https://doi.org/10.1016/j.dark.2023.101221}{\emph{Phys. Dark Univ.}
  {\bfseries 40} (2023) 101221}
  [\href{https://arxiv.org/abs/2203.07492}{{\ttfamily 2203.07492}}].

\bibitem{Essig:2011nj}
R.~Essig, J.~Mardon and T.~Volansky, \emph{{Direct Detection of Sub-GeV Dark
  Matter}}, \href{https://doi.org/10.1103/PhysRevD.85.076007}{\emph{Phys. Rev.}
  {\bfseries D85} (2012) 076007}
  [\href{https://arxiv.org/abs/1108.5383}{{\ttfamily 1108.5383}}].

\bibitem{Planck:2018vyg}
{\scshape Planck} collaboration, N.~Aghanim et~al., \emph{{Planck 2018 results.
  VI. Cosmological parameters}},
  \href{https://doi.org/10.1051/0004-6361/201833910}{\emph{Astron. Astrophys.}
  {\bfseries 641} (2020) A6}
  [\href{https://arxiv.org/abs/1807.06209}{{\ttfamily 1807.06209}}].

\bibitem{Lee:1977ua}
B.~W. Lee and S.~Weinberg, \emph{{Cosmological Lower Bound on Heavy Neutrino
  Masses}}, \href{https://doi.org/10.1103/PhysRevLett.39.165}{\emph{Phys. Rev.
  Lett.} {\bfseries 39} (1977) 165}.

\bibitem{LSND:2001akn}
{\scshape LSND} collaboration, L.~B. Auerbach et~al., \emph{{Measurement of
  electron - neutrino - electron elastic scattering}},
  \href{https://doi.org/10.1103/PhysRevD.63.112001}{\emph{Phys. Rev. D}
  {\bfseries 63} (2001) 112001}
  [\href{https://arxiv.org/abs/hep-ex/0101039}{{\ttfamily hep-ex/0101039}}].

\bibitem{MiniBooNEDM:2018cxm}
{\scshape MiniBooNE DM} collaboration, A.~A. Aguilar-Arevalo et~al.,
  \emph{{Dark Matter Search in Nucleon, Pion, and Electron Channels from a
  Proton Beam Dump with MiniBooNE}},
  \href{https://doi.org/10.1103/PhysRevD.98.112004}{\emph{Phys. Rev. D}
  {\bfseries 98} (2018) 112004}
  [\href{https://arxiv.org/abs/1807.06137}{{\ttfamily 1807.06137}}].

\bibitem{Akesson:2018vlm}
{\scshape LDMX} collaboration, T.~{\AA}kesson et~al., \emph{{Light Dark Matter
  eXperiment (LDMX)}},  \href{https://arxiv.org/abs/1808.05219}{{\ttfamily
  1808.05219}}.

\bibitem{Berlin_2019}
A.~Berlin, N.~Blinov, G.~Krnjaic, P.~Schuster and N.~Toro, \emph{Dark matter,
  millicharges, axion and scalar particles, gauge bosons, and other new physics
  with {LDMX}},
  \href{https://doi.org/10.1103/physrevd.99.075001}{\emph{Physical Review D}
  {\bfseries 99} (2019) }.

\bibitem{deNiverville_2019}
P.~deNiverville and C.~Frugiuele, \emph{Hunting sub-gev dark matter with the
  $\mathrm{NO}\ensuremath{\nu}\mathrm{A}$ near detector},
  \href{https://doi.org/10.1103/PhysRevD.99.051701}{\emph{Phys. Rev. D}
  {\bfseries 99} (2019) 051701}.

\bibitem{deNiverville_2017}
P.~deNiverville, C.-Y. Chen, M.~Pospelov and A.~Ritz, \emph{Light dark matter
  in neutrino beams: Production modeling and scattering signatures at
  {MiniBooNE}, t2k, and {SHiP}},
  \href{https://doi.org/10.1103/physrevd.95.035006}{\emph{Physical Review D}
  {\bfseries 95} (2017) }.

\bibitem{deNiverville:2016rqh}
P.~deNiverville, C.-Y. Chen, M.~Pospelov and A.~Ritz, \emph{{Light dark matter
  in neutrino beams: production modelling and scattering signatures at
  MiniBooNE, T2K and SHiP}},
  \href{https://doi.org/10.1103/PhysRevD.95.035006}{\emph{Phys. Rev. D}
  {\bfseries 95} (2017) 035006}
  [\href{https://arxiv.org/abs/1609.01770}{{\ttfamily 1609.01770}}].

\bibitem{E137}
B.~Batell, R.~Essig and Z.~Surujon, \emph{Strong constraints on sub-gev dark
  sectors from slac beam dump e137},
  \href{https://doi.org/10.1103/PhysRevLett.113.171802}{\emph{Phys. Rev. Lett.}
  {\bfseries 113} (2014) 171802}.

\bibitem{deFavereau:2013fsa}
{\scshape DELPHES 3} collaboration, J.~de~Favereau, C.~Delaere, P.~Demin,
  A.~Giammanco, V.~Lemaitre, A.~Mertens et~al., \emph{{DELPHES 3, A modular
  framework for fast simulation of a generic collider experiment}},
  \href{https://doi.org/10.1007/JHEP02(2014)057}{\emph{JHEP} {\bfseries 02}
  (2014) 057} [\href{https://arxiv.org/abs/1307.6346}{{\ttfamily 1307.6346}}].

\bibitem{deNiverville:2011it}
P.~deNiverville, M.~Pospelov and A.~Ritz, \emph{{Observing a light dark matter
  beam with neutrino experiments}},
  \href{https://doi.org/10.1103/PhysRevD.84.075020}{\emph{Phys. Rev. D}
  {\bfseries 84} (2011) 075020}
  [\href{https://arxiv.org/abs/1107.4580}{{\ttfamily 1107.4580}}].

\bibitem{Chu:2023zbo}
X.~Chu, J.~Hisano, A.~Ibarra, J.-L. Kuo and J.~Pradler, \emph{{Multipole vector
  dark matter below the GeV-scale}},
  \href{https://arxiv.org/abs/2303.13643}{{\ttfamily 2303.13643}}.

\bibitem{Gondolo:2021fqo}
P.~Gondolo, I.~Jeong, S.~Kang, S.~Scopel and G.~Tomar, \emph{{Phenomenology of
  nuclear scattering for a WIMP of arbitrary spin}},
  \href{https://doi.org/10.1103/PhysRevD.104.063018}{\emph{Phys. Rev. D}
  {\bfseries 104} (2021) 063018}
  [\href{https://arxiv.org/abs/2102.09778}{{\ttfamily 2102.09778}}].

\bibitem{Gondolo:2020wge}
P.~Gondolo, S.~Kang, S.~Scopel and G.~Tomar, \emph{{Effective theory of nuclear
  scattering for a WIMP of arbitrary spin}},
  \href{https://doi.org/10.1103/PhysRevD.104.063017}{\emph{Phys. Rev. D}
  {\bfseries 104} (2021) 063017}
  [\href{https://arxiv.org/abs/2008.05120}{{\ttfamily 2008.05120}}].

\bibitem{Catena:2019hzw}
R.~Catena, K.~Fridell and M.~B. Krauss, \emph{{Non-relativistic Eff`ective
  Interactions of Spin 1 Dark Matter}},
  \href{https://doi.org/10.1007/JHEP08(2019)030}{\emph{JHEP} {\bfseries 08}
  (2019) 030} [\href{https://arxiv.org/abs/1907.02910}{{\ttfamily
  1907.02910}}].

\bibitem{Catena:2018uae}
R.~Catena, K.~Fridell and V.~Zema, \emph{{Direct detection of fermionic and
  vector dark matter with polarised targets}},
  \href{https://doi.org/10.1088/1475-7516/2018/11/018}{\emph{JCAP} {\bfseries
  1811} (2018) 018} [\href{https://arxiv.org/abs/1810.01515}{{\ttfamily
  1810.01515}}].

\bibitem{Catena:2022fnk}
R.~Catena, D.~Cole, T.~Emken, M.~Matas, N.~Spaldin, W.~Tarantino et~al.,
  \emph{{Dark matter-electron interactions in materials beyond the dark photon
  model}}, \href{https://doi.org/10.1088/1475-7516/2023/03/052}{\emph{JCAP}
  {\bfseries 03} (2023) 052}
  [\href{https://arxiv.org/abs/2210.07305}{{\ttfamily 2210.07305}}].

\bibitem{Dent:2015zpa}
J.~B. Dent, L.~M. Krauss, J.~L. Newstead and S.~Sabharwal, \emph{{General
  analysis of direct dark matter detection: From microphysics to observational
  signatures}}, \href{https://doi.org/10.1103/PhysRevD.92.063515}{\emph{Phys.
  Rev.} {\bfseries D92} (2015) 063515}
  [\href{https://arxiv.org/abs/1505.03117}{{\ttfamily 1505.03117}}].

\bibitem{Choi:2019zeb}
S.-M. Choi, H.~M. Lee, Y.~Mambrini and M.~Pierre, \emph{{Vector SIMP dark
  matter with approximate custodial symmetry}},
  \href{https://doi.org/10.1007/JHEP07(2019)049}{\emph{JHEP} {\bfseries 07}
  (2019) 049} [\href{https://arxiv.org/abs/1904.04109}{{\ttfamily
  1904.04109}}].

\bibitem{LDMX}
T.~Åkesson, A.~Berlin, N.~Blinov, O.~Colegrove, G.~Collura, V.~Dutta et~al.,
  \emph{Light dark matter experiment (ldmx)},  2018.
\newblock 10.48550/ARXIV.1808.05219.

\bibitem{Catena_2018}
R.~Catena, J.~Conrad and M.~B. Krauss, \emph{Compatibility of a dark matter
  discovery at {XENONnT} or {LZ} with the {WIMP} thermal production mechanism},
  \href{https://doi.org/10.1103/physrevd.97.103002}{\emph{Physical Review D}
  {\bfseries 97} (2018) }.

\bibitem{Arcadi:2017kky}
G.~Arcadi, M.~Dutra, P.~Ghosh, M.~Lindner, Y.~Mambrini, M.~Pierre et~al.,
  \emph{{The waning of the WIMP? A review of models, searches, and
  constraints}},
  \href{https://doi.org/10.1140/epjc/s10052-018-5662-y}{\emph{Eur. Phys. J.}
  {\bfseries C78} (2018) 203}
  [\href{https://arxiv.org/abs/1703.07364}{{\ttfamily 1703.07364}}].

\bibitem{Berlin:2018bsc}
A.~Berlin, N.~Blinov, G.~Krnjaic, P.~Schuster and N.~Toro, \emph{{Dark Matter,
  Millicharges, Axion and Scalar Particles, Gauge Bosons, and Other New Physics
  with LDMX}},  \href{https://arxiv.org/abs/1807.01730}{{\ttfamily
  1807.01730}}.

\bibitem{PhysRevD.102.095011}
A.~Berlin, P.~deNiverville, A.~Ritz, P.~Schuster and N.~Toro, \emph{Sub-gev
  dark matter production at fixed-target experiments},
  \href{https://doi.org/10.1103/PhysRevD.102.095011}{\emph{Phys. Rev. D}
  {\bfseries 102} (2020) 095011}.

\bibitem{Alloul:2013bka}
A.~Alloul, N.~D. Christensen, C.~Degrande, C.~Duhr and B.~Fuks,
  \emph{{FeynRules 2.0 - A complete toolbox for tree-level phenomenology}},
  \href{https://doi.org/10.1016/j.cpc.2014.04.012}{\emph{Comput. Phys. Commun.}
  {\bfseries 185} (2014) 2250}
  [\href{https://arxiv.org/abs/1310.1921}{{\ttfamily 1310.1921}}].

\bibitem{Belyaev:2012qa}
A.~Belyaev, N.~D. Christensen and A.~Pukhov, \emph{{CalcHEP 3.4 for collider
  physics within and beyond the Standard Model}},
  \href{https://doi.org/10.1016/j.cpc.2013.01.014}{\emph{Comput. Phys. Commun.}
  {\bfseries 184} (2013) 1729}
  [\href{https://arxiv.org/abs/1207.6082}{{\ttfamily 1207.6082}}].

\bibitem{Belanger:2006is}
G.~Belanger, F.~Boudjema, A.~Pukhov and A.~Semenov, \emph{{MicrOMEGAs 2.0: A
  Program to calculate the relic density of dark matter in a generic model}},
  \href{https://doi.org/10.1016/j.cpc.2006.11.008}{\emph{Comput. Phys. Commun.}
  {\bfseries 176} (2007) 367}
  [\href{https://arxiv.org/abs/hep-ph/0607059}{{\ttfamily hep-ph/0607059}}].

\bibitem{Gondolo:1990dk}
P.~Gondolo and G.~Gelmini, \emph{{Cosmic abundances of stable particles:
  Improved analysis}},
  \href{https://doi.org/10.1016/0550-3213(91)90438-4}{\emph{Nucl. Phys.}
  {\bfseries B360} (1991) 145}.

\bibitem{Izaguirre_2016}
E.~Izaguirre, G.~Krnjaic and B.~Shuve, \emph{Discovering inelastic thermal
  relic dark matter at colliders},
  \href{https://doi.org/10.1103/physrevd.93.063523}{\emph{Physical Review D}
  {\bfseries 93} (2016) }.

\bibitem{PDG}
Z.~et~al. (Particle Data~Group), \emph{{Review of Particle Physics}},
  \href{https://doi.org/10.1093/ptep/ptaa104}{\emph{Progress of Theoretical and
  Experimental Physics} {\bfseries 2020} (2020) }
  [\href{https://arxiv.org/abs/https://academic.oup.com/ptep/article-pdf/2020/8/083C01/34673722/ptaa104.pdf}{{\ttfamily
  https://academic.oup.com/ptep/article-pdf/2020/8/083C01/34673722/ptaa104.pdf}}].

\bibitem{serendipity}
P.~Ilten, Y.~Soreq, M.~Williams and W.~Xue, \emph{Serendipity in dark photon
  searches}, \href{https://doi.org/10.1007/jhep06(2018)004}{\emph{Journal of
  High Energy Physics} {\bfseries 2018} (2018) }.

\bibitem{Catena_2023}
R.~Catena, D.~Cole, T.~Emken, M.~Matas, N.~Spaldin, W.~Tarantino et~al.,
  \emph{Dark matter-electron interactions in materials beyond the dark photon
  model}, \href{https://doi.org/10.1088/1475-7516/2023/03/052}{\emph{Journal of
  Cosmology and Astroparticle Physics} {\bfseries 2023} (2023) 052}.

\bibitem{CMB_swave}
T.~R. Slatyer, \emph{Indirect dark matter signatures in the cosmic dark ages.
  i. generalizing the bound on $s$-wave dark matter annihilation from planck
  results}, \href{https://doi.org/10.1103/PhysRevD.93.023527}{\emph{Phys. Rev.
  D} {\bfseries 93} (2016) 023527}.

\bibitem{MeVDMComplementarity}
M.~Dutra, M.~Lindner, S.~Profumo, F.~S. Queiroz, W.~Rodejohann and C.~Siqueira,
  \emph{{MeV} dark matter complementarity and the dark photon portal},
  \href{https://doi.org/10.1088/1475-7516/2018/03/037}{\emph{Journal of
  Cosmology and Astroparticle Physics} {\bfseries 2018} (2018) 037}.

\bibitem{Liu_2016}
H.~Liu, T.~R. Slatyer and J.~Zavala, \emph{Contributions to cosmic reionization
  from dark matter annihilation and decay},
  \href{https://doi.org/10.1103/physrevd.94.063507}{\emph{Physical Review D}
  {\bfseries 94} (2016) }.

\bibitem{Liu_2021}
H.~Liu, W.~Qin, G.~W. Ridgway and T.~R. Slatyer,
  \emph{Lyman-$\ensuremath{\alpha}$ constraints on cosmic heating from dark
  matter annihilation and decay},
  \href{https://doi.org/10.1103/PhysRevD.104.043514}{\emph{Phys. Rev. D}
  {\bfseries 104} (2021) 043514}.

\bibitem{NA64:2019imj}
D.~Banerjee et~al., \emph{{Dark matter search in missing energy events with
  NA64}}, \href{https://doi.org/10.1103/PhysRevLett.123.121801}{\emph{Phys.
  Rev. Lett.} {\bfseries 123} (2019) 121801}
  [\href{https://arxiv.org/abs/1906.00176}{{\ttfamily 1906.00176}}].

\bibitem{Schuster:2021mlr}
P.~Schuster, N.~Toro and K.~Zhou, \emph{{Probing invisible vector meson decays
  with the NA64 and LDMX experiments}},
  \href{https://doi.org/10.1103/PhysRevD.105.035036}{\emph{Phys. Rev. D}
  {\bfseries 105} (2022) 035036}
  [\href{https://arxiv.org/abs/2112.02104}{{\ttfamily 2112.02104}}].

\bibitem{Lees:2017lec}
{\scshape BaBar} collaboration, J.~P. Lees et~al., \emph{{Search for Invisible
  Decays of a Dark Photon Produced in ${e}^{+}{e}^{-}$ Collisions at BaBar}},
  \href{https://doi.org/10.1103/PhysRevLett.119.131804}{\emph{Phys. Rev. Lett.}
  {\bfseries 119} (2017) 131804}
  [\href{https://arxiv.org/abs/1702.03327}{{\ttfamily 1702.03327}}].

\bibitem{belle2}
E.~Kou et~al., \emph{{The Belle II Physics Book}},
  \href{https://doi.org/10.1093/ptep/ptz106}{\emph{Progress of Theoretical and
  Experimental Physics} {\bfseries 2019} (2019) }
  [\href{https://arxiv.org/abs/https://academic.oup.com/ptep/article-pdf/2019/12/123C01/32693980/ptz106.pdf}{{\ttfamily
  https://academic.oup.com/ptep/article-pdf/2019/12/123C01/32693980/ptz106.pdf}}].

\bibitem{Kahlhoefer:2015bea}
F.~Kahlhoefer, K.~Schmidt-Hoberg, T.~Schwetz and S.~Vogl, \emph{{Implications
  of unitarity and gauge invariance for simplified dark matter models}},
  \href{https://doi.org/10.1007/JHEP02(2016)016}{\emph{JHEP} {\bfseries 02}
  (2016) 016} [\href{https://arxiv.org/abs/1510.02110}{{\ttfamily
  1510.02110}}].

\bibitem{Chang:2023cki}
C.~Chang, P.~Scott, T.~E. Gonzalo, F.~Kahlhoefer and M.~White, \emph{{Global
  fits of simplified models for dark matter with GAMBIT II. Vector dark matter
  with an $s$-channel vector mediator}},
  \href{https://arxiv.org/abs/2303.08351}{{\ttfamily 2303.08351}}.

\end{thebibliography}\endgroup

\end{document}